\documentclass[11pt]{article}

\usepackage{amssymb}
\usepackage{amsmath}
\usepackage{graphicx}
\usepackage[cm]{fullpage}
\usepackage{booktabs}
\usepackage{siunitx}
\usepackage{lineno}
\usepackage{hyperref}
\usepackage[noabbrev]{cleveref}
\usepackage{authblk}
\usepackage{natbib}

\DeclareSIUnit{\degree}{\text{deg}}
\DeclareSIUnit{\revolution}{\text{rev}}

\usepackage{nomencl}
\makenomenclature
\usepackage{etoolbox}
\renewcommand\nomgroup[1]{%
    \item[\bfseries
    \ifstrequal{#1}{a}{Roman}{%
    \ifstrequal{#1}{b}{Roman Vectors}{%
    \ifstrequal{#1}{c}{Greek}{%
    \ifstrequal{#1}{d}{Acronyms}{}
    }}}]}

\graphicspath{{./img/}}
\title{Model-scale experiments of passive pitch control for tidal turbines}

\author[1]{Stefano Gambuzza}
\author[1]{Gabriele Pisetta}
\author[2]{Thomas Davey}
\author[2]{Jeffrey Steynor}
\author[1,$\dagger$]{Ignazio Maria Viola}
\affil[1]{School of Engineering, Institute for Energy Systems, University of Edinburgh EH9 3FB Edinburgh, United Kingdom}
\affil[2]{FloWave Ocean Energy Research Facility, University of Edinburgh EH9 3FB Edinburgh, United Kingdom}
\affil[$\dagger$]{Corresponding author, i.m.viola@ed.ac.uk}

\date{}

\begin{document}
\maketitle


\begin{abstract}
Tidal currents are renewable and predictable energy sources that could prove fundamental to decrease dependency from fossil fuels.
Tidal currents, however, are highly unsteady and non uniform, resulting in undesirable load fluctuations on the blades and the drive train of turbines.
A passive morphing blade concept capable to reduce the load fluctuations without affecting the mean loads has recently been formulated and demonstrated with numerical simulations \citep{Pisetta2022}.
In this paper, we present the first demonstration of this morphing blade concept, through experimental tests on a 1.2-m diameter turbine. We show that fluctuations in the root-bending moment, thrust and torque are consistently reduced over a broad range of tip-speed ratios.
This work also highlights some critical design aspects of morphing blades.
For instance, it is showed that the friction resistance can substantially decrease the effectiveness of the system and thus must be minimised by design. Overall this paper demonstrates for the first time the effectiveness of morphing blades for tidal turbines, paving the way to the future development of this technology.
\end{abstract}



\nomenclature[a]{$Q$}{Torque generated by the turbine [\si{\newton\meter}]}
\nomenclature[a]{$T$}{Thrust generated by the turbine [\si{\newton}]}
\nomenclature[a]{$M_y$}{Root-bending moment generated by each blade [\si{\newton\meter}]}
\nomenclature[a]{$C_Q$}{Torque coefficient}
\nomenclature[a]{$C_T$}{Thrust coefficient}
\nomenclature[a]{$C_y$}{Root-bending moment coefficient}
\nomenclature[a]{$S_Q$}{Spectrum of torque fluctuations [\si{\newton\squared\meter\squared\per\hertz}]}
\nomenclature[a]{$S_T$}{Spectrum of thrust fluctuations [\si{\newton\squared\per\hertz}]}
\nomenclature[a]{$S_y$}{Spectrum of root-bending moment fluctuations [\si{\newton\squared\meter\squared\per\hertz}]}
\nomenclature[a]{$G_Q$}{Transfer function of torque fluctuations}
\nomenclature[a]{$G_T$}{Transfer function of thrust fluctuations}
\nomenclature[a]{$G_y$}{Transfer function of root-bending moment fluctuations}
\nomenclature[a]{$J$}{Rotor moment of inertia [\si{\kilo\gram\meter\squared}]}
\nomenclature[a]{$k$}{Spring elastic constant [\si{\newton\metre\per\degree}]}
\nomenclature[a]{$M_s$}{Spring elastic moment [\si{\newton\metre}]}
\nomenclature[a]{$M_h$}{Hydrodynamic pitching moment [\si{\newton\metre}]}
\nomenclature[a]{$M_\mathrm{fric}$}{Moment due to friction around the pitching axis [\si{\newton\metre}]}
\nomenclature[a]{$z$}{Height from facility bed [\si{\meter}]}
\nomenclature[a]{$z_\mathrm{hub}$}{Hub-height measured from facility bed [\si{\meter}]}
\nomenclature[a]{$r$}{Coordinate along the blade spanwise direction [\si{\meter}]}
\nomenclature[a]{$c$}{Local chord of the turbine blade [\si{\meter}]}
\nomenclature[a]{$D$}{Turbine rotor diameter [\si{\meter}]}
\nomenclature[a]{$x_P$}{Distance of the pitching axis from the blade leading edge [\si{\meter}]}
\nomenclature[a]{$t$}{Local blade thickness [\si{\meter}]}
\nomenclature[a]{$L$}{Lift generated by a section of the blade [\si{\newton}]}
\nomenclature[a]{$C_l$}{Coefficient of lift generated by a section of the blade}
\nomenclature[a]{$U$}{Flow velocity in the streamwise direction [\si{\meter\per\second}]}
\nomenclature[a]{$U_\infty$}{Free current velocity [\si{\meter\per\second}]}
\nomenclature[a]{$S_u$}{Spectrum of streamwise velocity fluctuations [\si{\meter\squared\per\second}]}
\nomenclature[a]{$Re_D$}{Diameter-based Reynolds number}
\nomenclature[a]{$I_\infty$}{Free current turbulence intensity}


\nomenclature[c]{$\Delta Q'$}{Relative difference in torque fluctuations}
\nomenclature[c]{$\Delta T'$}{Relative difference in thrust fluctuations}
\nomenclature[c]{$\Delta M_y'$}{Relative difference in root-bending moment fluctuations}
\nomenclature[c]{$\omega$}{Turbine angular velocity [\si{\revolution\per\min}]}
\nomenclature[c]{$\lambda$}{Tip-speed ratio}
\nomenclature[c]{$G_\omega$}{Transfer function of angular velocity fluctuations}
\nomenclature[c]{$S_\omega$}{Spectrum of velocity fluctuations [\si{\revolution\squared\per\min\squared\per\hertz}]}
\nomenclature[c]{$\Delta \omega'$}{Relative difference in angular velocity fluctuations}
\nomenclature[c]{$\beta_0$}{Local blade twist [\si{\degree}]}
\nomenclature[c]{$\beta$}{Instantaneous blade pitch [\si{\degree}]}
\nomenclature[c]{$\beta_\mathrm{pre}$}{Spring preload angle [\si{\degree}]}
\nomenclature[c]{$\rho$}{Water density [\si{\kilo\gram\per\meter\cubed}]}
\nomenclature[c]{$\theta$}{Water temperature [$^\circ\mathrm{C}$]}
\nomenclature[c]{$\nu$}{Water kinematic viscosity [\si{\meter\squared\per\second}]}
\nomenclature[c]{$\mu$}{Water dynamic viscosity [\si{\kilo\gram\per\meter\per\second}]}
\nomenclature[c]{$\alpha$}{Angle of attack of a section of the blade [\si{\degree}]}
\nomenclature[c]{$\alpha_\mathrm{sh}$}{Shear profile coefficient}

\printnomenclature

\section{Introduction} \label{sec:intro}
Tidal energy is a renewable energy source that has the potential to provide a large share of usable power to the national grid in a predictable fashion \citep{Neill2017, Taveira-Pinto2020}.
In the last couple of years, the largest tidal energy generator, the 2 MW Orbital Marine Power’s O2, has commenced generation at the European Marine Energy Centre in Orkney.
The world’s first tidal array by Nova Innovation in Shetland has been expended demonstrating the first electric vehicle charging point powered entirely from a tidal energy source.
The largest planned tidal project, MeyGen, owned and operated by SIMEC Atlantis Energy in the Scotland's Pentland Firth, is being deployed and has the capacity to install 100 MW by 2024.
However, despite the fast growth of the sector, there is a pressing need to decrease the levelised cost of energy to become competitive with the other renewable energy sectors \citep{SupergenORE2021, PolicyandInnovationGroup2021}.

Tidal turbines experience high load fluctuations due to the high turbulence intensity and the shear of the tidal stream, wave-induced currents \citep{Scarlett2019, Scarlett2020, Adcock2021}, and the interaction with neighbouring turbines \citep{Vogel2019}. 
Load fluctuations translate to an increased cost of tidal energy as they reduce the mean-time-to-failure of turbine blades and drive trains \citep{Chen2015}, as well as requiring over-dimensioned generators and support structures able to withstand the load peaks instead of the maximum time-mean loads.
The changes in loading on the turbine can be significant: model-scale experiments in a towing tank show an increase in the root-bending moment in the range of \SIrange{15}{25}{\percent} with respect to its mean value \citep{Milne2015, Milne2016}, while tests in wave tanks show an increase of more than \SI{100}{\percent} in the worst-case conditions \citep{Galloway2014}, with obvious effects on the structural integrity of the blades.
The effects of the unsteady loads also influence on the thrust generated by the turbines: \cite{Barltrop2006} shows measurements of thrust for a tidal turbine subject to wakes of height comparable to the blade length, highlighting that peak thrust is \SI{40}{\percent} larger than the mean value.
Similarly, \cite{Ahmed2017} and \cite{Parkinson2016} show comparable increases both for power and thrust for a tidal turbine operating at the EMEC site in the Orkney Islands, Scotland.

To reduce the magnitude of these effects, collective pitch control is often implemented on megawatt-scale turbines.
\citet{Kennedy2018} showed that blades of turbines equipped with collective pitch control need \SI{10}{\percent} thinner laminates than blades of equivalent stall-regulated turbines.
Actively controlled trailing edge flaps have not yet been deployed on commercial-scale turbines, but \citet{Bernhammer2016} estimated that these could reduce fatigue loads by \SI{59}{\percent} in the best-case scenario for a flap that spans \SI{30}{\percent} of the blade span and \SI{10}{\percent} of its chord.

Unfortunately, active control systems for tidal turbines are a driver in the operating costs of tidal turbines because of the potential for failures and the need for maintenance \citep{Johnstone2013}.
For this reason, a number of studies have investigated the substitution of active control systems with passive systems requiring less maintenance.
Note that, in these studies, passive control does not refer uniquely to stall-control, as it is customary for wind turbines \citep{Balat2009}, but instead it refers to more complex systems capable of passively actuating pitching motions in response to instantaneous loads.
Both \citet{Bottasso2016} and \citet{Cordes2018} have introduced passive mechanisms to actuate either a trailing-edge flap or a change in the leading edge camber in response to the instantaneous loading on the blades.
While these devices appear promising, they still require complex linkage systems and bearings that are exposed to debris and biofouling \citep{Stringer2020}.
Multiple studies have proposed mechanisms to change the geometry of the whole blade in response to its instantaneous loading, either by allowing it to passively pitch around a given axis or by exploiting the elastic properties of composite materials.
For instance, \citet{Cheney1978} proposed a mechanism where the blade is free to pitch around an axis along the span-wise direction, and it is connected to an eccentric mass; the moment generated by the centrifugal force acting on the mass around the pitching axis pitches the whole blade to feather when the rotor speed increases, thus reducing the loads generated and reducing the rotor angular velocity.
\cite{Karaolis1988} introduced the idea of biased composite lay-ups to achieve twist coupling in response to blade bending or centrifugal loads: with increasing wind speed the blade would pitch to stall, therefore regulating power generation. The concept gained popularity as a promising system to improve energy yield and torque startup \citep{Kooijman1996} but it has seen limited application due to the difficulty of predicting power in post-stall regimes and manufacturing limitations \citep{Veers1998}.
A more recent example are the flexible blades designed by \citet{Cognet2020}, where the blade flexibility was optimised to increase the extracted power by \SI{35}{\percent} for wind conditions typical of the North Sea.
Several other designs leveraging on the flexibility of the composite materials of the blades have been tested over the years, both for wind turbines \citep{Pavese2017} and for tidal machines \citep{Murray2016, Murray2018, Porter2020}.
In most cases, the main objective of these mechanisms is the possibility of passively limiting the power transferred to the turbine generator across a range of freestream speeds \citep{Krawczyk2013}, and not that of increasing the blades life-span by reducing the impact of fatigue loading.

\citet{Viola2021,Viola2022p} analysed theoretically the underlying mechanisms of passive unsteady loading alleviation, including the deformation of a flexible trailing edge, the passive actuation of a flap, or the passive pitch of a rigid blade.
This study suggests that the fraction of unsteady load mitigation is equal to the ratio between the flap's chord and the blade's chord.
Therefore, allowing for the whole blade to passively pitch, instead of restricting this motion to the trailing edge such as the case of a passive flap or flexible trailing edge, can result in a full removal of the unsteady loads.
This theoretical framework was further investigated by \citet{Pisetta2022}, who developed a numerical code based on blade element momentum theory and Theodorsen's theory, demonstrating the unsteady load mitigation potential.
Both the theoretical work of \citet{Viola2021,Viola2022p} and the low-order modelling of \citet{Pisetta2022} are relevant to all three morphing blades concepts: flexible trailing edge, trailing edge flap, and passive pitch of a rigid blade.
The passive trailing edge flap concept was further investigated through water tunnels tests of an extruded blade sections by \cite{Arredondo-Galeana2021}.
They verified that a flap with a hinge at 25\% of the chord allows a 25\% reduction of the unsteady loading.
These results further motivated the numerical studies of \citet{Dai2022} on the passive pitch of rigid blades, a mechanism that could theoretically lead to the total cancellation of the unsteady loads.
Using computational fluid dynamics simulations of a turbine in a sheared current, the authors found that a turbine equipped with passively pitching blades experiences a reduction in the unsteady thrust of approximately \SI{80}{\percent}, with no effect on its time-average value.
The authors however note that their results have neglected the effects of friction on the pitching motion on the turbine, and they stress the need to verify whether this assumption is practically achievable.
The present paper is based on the previous works of \citet{Viola2021,Viola2022p, Pisetta2022} and \cite{Dai2022}, and aims to demonstrate with physical experiments on a model scale turbine the unsteady load reduction potential of passively pitching rigid blades.
Measurements of the instantaneous torque, thrust, and root-bending moment are provided for a range of tip speed ratios, and the effect of friction on the pitching motion is carefully analysed.

The paper is structured as follows: \cref{sec:method} outlines the experimental methodology, including a description of the facility, the geometry of the model-scale turbine and the implementation of the passive pitch mechanism.
\Cref{sec:results} presents the experimental results, assessing the effectiveness of the passive pitch design by comparing the mean and time-varying loads on the morphing blade to those experienced by a rigid blade having the same geometry; attention will be dedicated to the technical challenges that have the potential to reduce the effectiveness of the passive pitch mechanism.
\Cref{sec:conclusions} will summarise these findings in a concise manner.

\section{Methodology}\label{sec:method}
The experimental campaign has been carried out in the University of Edinburgh's FloWave Ocean Energy Research Facility (hereinafter, FloWave), using a model-scale tidal turbine.
These experiments are presented in this section, along with details on the instrumentation used to acquire the data presented in this study and limitations on the experimental techniques used.

\subsection{Facility and turbine geometry} \label{ssec:facility}
\begin{figure}[ht!]
    \centering
    \includegraphics[scale=0.5]{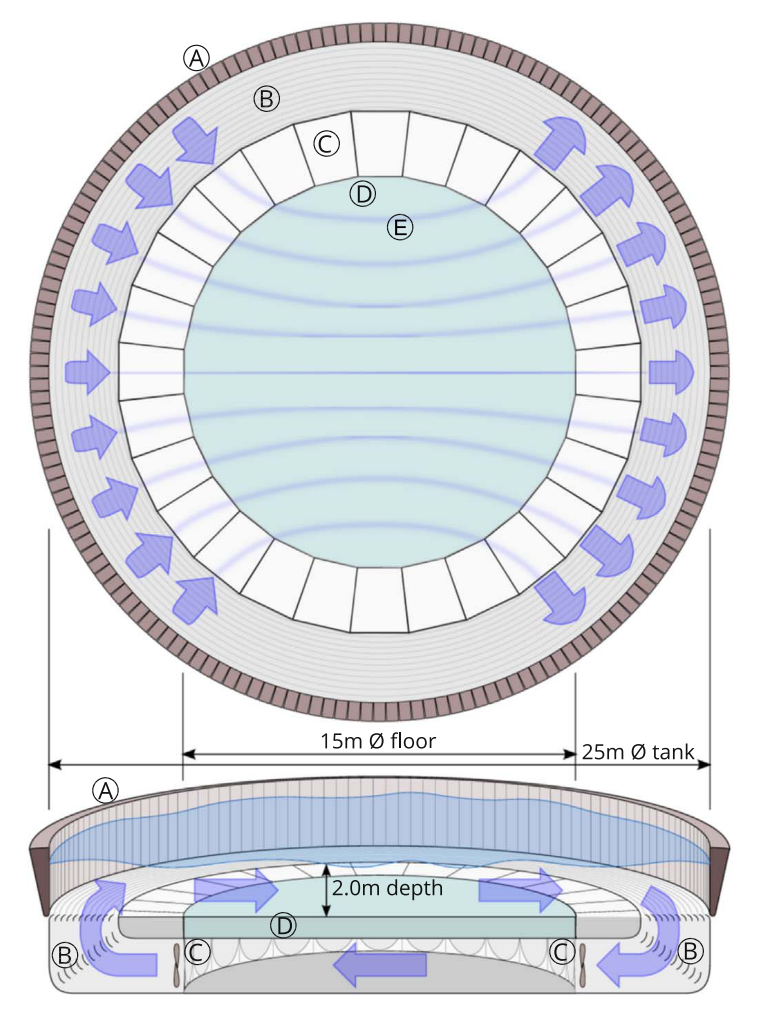}
    \caption{Schematic of FloWave in plan and oblique section showing: A) Wavemaker paddles around circumference; B) Turning vanes and flow conditioning filters; C) Current drive impeller units; D) Buoyant raisable floor below test area; E) Idealised streamlines of flow across tank floor.}
    \label{fig:flowave}
\end{figure}

The FloWave facility has a cylindrical test section with a diameter of \SI{25}{\meter} and a water depth of \SI{2}{\meter}.
The facility was operated to generate a current velocity having a hub-height speed $U_\infty =  \SI{0.8}{\meter\per\second}$, which was kept constant between test cases.
The vertical velocity profile followed a canonical power law
\begin{equation}
    \frac{U(z)}{U_\infty} = \left(\frac{z}{z_\mathrm{hub}}\right)^{\alpha_\mathrm{sh}},
\end{equation}
where $z$ is the vertical direction with origin on the test section floor, $z_\mathrm{hub} = \SI{876}{\milli\meter}$ is the turbine hub-height from the test section bed, and $\alpha_\mathrm{sh} = 1/15$ for this facility \citep{Noble2015}.
At hub-height, the free-stream turbulence intensity $I_\infty$ was \SI{7.5}{\percent}.
This is defined as
\begin{equation}
    I_\infty = \frac{\sqrt{\overline{u'^2}}}{U_\infty},
\end{equation}
where $u'$ is the fluctuating component of the velocity and the overline represents time-averaging, so that $\sqrt{\overline{u'^2}}$ is the standard deviation of the velocity.
A schematic view of the test section is reported in \cref{fig:flowave}.

The model-scale tidal turbine used in this experimental campaign is a 1:15 scale model of the Tidal Generation Ltd. (now SABELLA) \SI{1}{\mega\watt} tidal turbine, and is thus representative of a typical tidal turbine geometry.
The drivetrain is the same used by, for instance, \cite{Payne2017} and \cite{Viola2022}.
It has a rotor diameter $D$ of \SI{1.2}{\meter} with the blades mounted on a hub of diameter $D_\mathrm{hub}$ of \SI{0.12}{\meter}; this results in a diameter-based Reynolds number $Re_D$ of \num{9.8e5}, defined as
\begin{equation}
    Re_D = \frac{U_\infty \, D}{\nu},
\end{equation}
where $\nu$ is the kinematic viscosity of water.
The turbine rotor is driven by a brushless permanent magnet servo motor, to which the turbine rotor was directly connected.
The motor was operated to drive the turbine rotor to a mandated angular velocity $\omega$, and thus a fixed tip-speed ratio
\begin{equation}
    \lambda\ = \frac{\omega D/2}{U_\infty}.
\end{equation}
Further information on the drive system is reported in \citet[sec.~6]{Payne2017}.

The turbine blades were manufactured by Piran Advanced Composites from \SI{3}{\milli\meter} thick carbon-fibre; these are filled with water during operation, which reduces the periodic loads generated due to the blades buoyancy as these rotate around the turbine axis.
The full blade table, comprising the chord, twist and thickness distributions along the spanwise direction, is reported by \citet{Gretton2010} and \citet{Pisetta2022a}.
The blade is able to pitch rigidly around a pitching axis, which is directed in the span-wise direction; each section of the blade is positioned so that the pitching axis passes through the section chord.
The distance between the leading edge of each section and the pitching axis is reported in \cref{tab:bladetable} as $x_P/c$: a value comprised between \numlist{0;1} denotes that the pitching axis crosses the profile chord between the leading and the trailing edge.
With the exception of the third of the blade closest to the root, all profiles pitch around a point which is at the tenth of their chord.
Apart from section \num{1}, which is circular, all sections are aerofoils of the \mbox{NACA 63$_{(3)}$-4XX} family, themselves a subset of the NACA $6$-Series aerofoils.
The numbering system for this aerofoil is as follows.
The second digit ($3$) denotes the location of the minimum pressure in tenths of the chord.
The fourth digit ($4$) indicates the ideal lift coefficient $C_l$ at a zero angle of attack, in tenths.
The third digit ($_{(3)}$) represents the half-width of the range of $C_l$ around this ideal value for which a favourable pressure gradient exists around the aerofoil, in tenths.
Lastly, the last two digits represent the maximum thickness in hundredths of the chord.
For a more thorough definition of the numbering system, the reader is referred to \citet[\S 6.8.c]{Abbott1959}.

\begin{table}[ht!]
    \centering
	\caption{Location of the pitching axis $x_P/c$ along the spanwise (radial) direction.}
    \label{tab:bladetable}
    \vspace{6pt}
    \begin{tabular}{lrr} \toprule
         Section & $r/D$ & $x_P/c$ \\ \midrule
		 1       & 0.092 & 0.50    \\
		 2       & 0.117 & 0.32    \\
		 3       & 0.133 & 0.20    \\
		 4       & 0.158 & 0.15    \\
		 5       & 0.183 & 0.10    \\
		 6       & 0.200 & 0.10    \\
		 7       & 0.225 & 0.10    \\
		 8       & 0.250 & 0.10    \\
		 9       & 0.267 & 0.10    \\
		 10      & 0.292 & 0.10    \\
		 11      & 0.317 & 0.10    \\
		 12      & 0.333 & 0.10    \\
		 13      & 0.358 & 0.10    \\
		 14      & 0.392 & 0.10    \\
		 15      & 0.400 & 0.10    \\
		 16      & 0.425 & 0.10    \\
		 17      & 0.450 & 0.10    \\
		 18      & 0.467 & 0.10    \\
		 19      & 0.492 & 0.10    \\
		 20      & 0.500 & 0.10    \\ \bottomrule
    \end{tabular}
\end{table}

\subsection{Passive pitch mechanism} \label{ssec:ppitch_mech}
\begin{figure}[ht!]
    \centering
    \includegraphics[width=0.75\columnwidth]{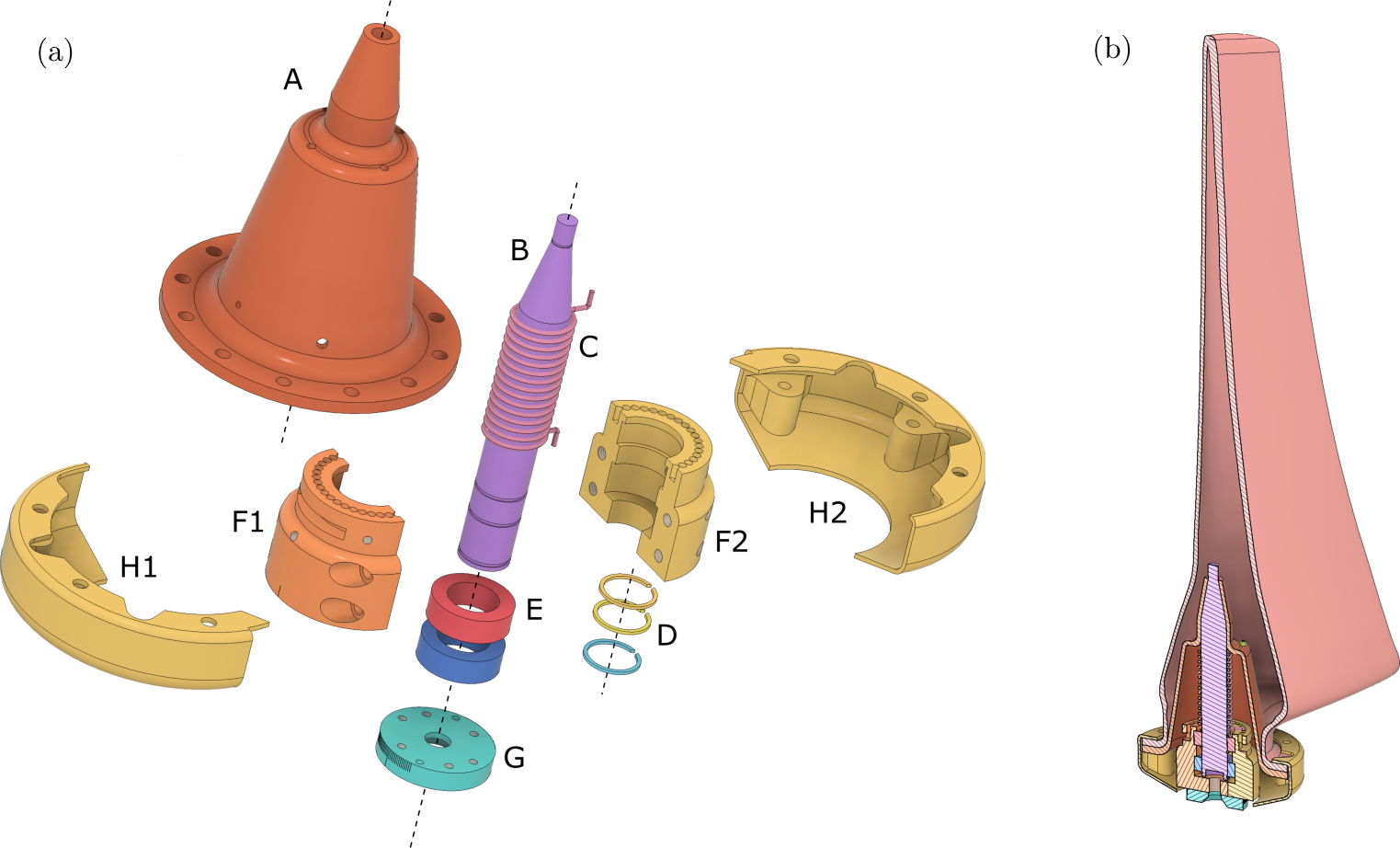}
    \caption{(a) Exploded view of the passive pitching mechanism: A) flanged cone; B) shaft; C) torsional spring; D) circlips to retain E) bearings; F1), F2) casing halves; G) angular
plate; H1), H2) fairing; (b) Section view of the assembled passive pitch mechanism.}
    \label{fig:pitch-exploded}
\end{figure}

The passive pitching mechanism is realised by means of a torsional spring wound on the pitching shaft: an exploded view presenting all components is provided in \cref{fig:pitch-exploded}(a), while the assembled system is presented in \cref{fig:pitch-exploded}(b) in sectioned view.
The blade shell is rigidly mounted on the inner cone via the common flange, and the pitching shaft is glued to the cone by means of epoxy resin.
The system is supported by two rolling-element bearings, which allow for the system to rotate around the pitching axis.
This rotation is constrained by the torsion spring, which is engaged both to the turbine frame via the casing halves and to the blade via holes on the flanged cone.
The bearing casing is then enshrouded in a 3D-printed hydrodynamic fairing.
This mechanism is then mounted on the model-scale turbine hub by means of an intermediate component labelled root-bending flexure, which is instrumented with strain gauges to measure the root-bending moment; this component, not pictured in \cref{fig:pitch-exploded}, sits below the angular plate (G).
Mounting between the angular plate and the casing halves is obtained by means of bolts that engage threaded holes in the casing halves, and pass through clearance holes in the angular plate; the necessary clearance between the holes and the bolts results in play that affects the geometrical pitching angle of the whole system, adding a degree of uncertainty to the actual pitching of the turbine blades.
Moreover, it was found out during the first tests that one of the blades featured a different pitch angle due to the small inaccuracy of a manufactured part: this blade is named Blade 2 in the remainder of the text.

To provide an estimate of the optimal spring stiffness that reduces the unsteadiness in the root-bending moment, the low-order code of \citet{Pisetta2022} has been used to simulate the loads on the blade when operating at $\lambda=7$ and at $U_\infty= \SI{0.8}{\metre\per\second}$.
The spring used during these tests has a stiffness $k = \SI{4.2}{\newton\milli\meter\per\degree}$; the spring is preloaded by an angle $\beta_\mathrm{pre}$, and the moment it generates around the pitching axis is
\begin{equation}
    M_s(\beta) = k \left(\beta + \beta_\mathrm{pre}\right),
    \label{eq:spring-moment}
\end{equation}
where $\beta$ is the instantaneous pitch angle, zero at on-design conditions, so that \mbox{$M_s(0) = k \, \beta_\mathrm{pre}$} balances the hydrodynamic moment of the blade at the on-design conditions and the blade is in equilibrium around this position.
The passive pitching system is designed so that an increase in the loading on the blade increases the pitching moment to be larger than the torque provided by the spring, and the blade rigidly pitches upstream (nose-down) to decrease the angle of attack experienced by each individual section; this in turn reduces the blade loading and the hydrodynamic moment it generates until a new equilibrium is found.
Vice versa, the opposite holds for an instantaneous decrease in the blade loading, so that the equilibrium around $\beta = 0$ is stable.
The full dynamics of the blade can thus be schematised by the following differential equation:

\begin{equation}
    J \ddot{\beta} = -M_s(\beta) + M_h(\beta, \dot{\beta}) + M_\mathrm{fric}(\dot{\beta}),
    \label{eq:ode_pitching}
\end{equation}
where $J$ is the moment of inertia of the blade around the pitching axis, $M_h$ is the hydrodynamic pitching moment generated by the integral of the moment generated by the fluid's pressure and shear fields on the turbine blade around the pitching axis, and $M_\mathrm{fric}$ includes the friction generated by the bearings and the spring around the blade.

At steady-state, one has that $\ddot{\beta} = \dot{\beta} = 0$, and the average pitching angle of the blades is given by
\begin{equation}
    M_s(\beta = 0) = M_h(\beta = 0) + M_\mathrm{fric}(\dot{\beta}=0),
    \label{eq:ss_pitching}
\end{equation}
where the last term is the moment due to static friction.
With the definition of the spring moment given in \cref{eq:spring-moment}, \cref{eq:ss_pitching} can be solved for $\beta_\mathrm{pre}$ to yield
\begin{equation}
    \beta_\mathrm{pre} = \frac{1}{k} (M_h(\beta = 0) + M_\mathrm{fric}(\dot{\beta} = 0)).
    \label{eq:betapre}
\end{equation}
The estimation of the preload angle therefore assumes knowledge of both the hydrodynamic pitching moment and the static friction around the pitching shaft.
The former can be estimated accurately with, for instance, a blade-element momentum algorithm, and the latter often relies on simplified engineering models.
Note that the sign of $M_\mathrm{fric}(\dot{\beta}=0)$ must be opposite to that of $M_h(\beta=0)$, as this opposes the motion of the blade due to the resultant of the hydrodynamic pressure.
In practical applications this suggests that the steady-state performance of the passive pitch system can be tuned to the presence of friction by reducing the spring preload from the value computed without friction.

The choice of preload angle is therefore dictated by the value of the spring stiffness chosen, as the main constraint on this is that the product $k \, \beta_\mathrm{pre}$ be equal to the pitching moment generated by the blade in operating conditions.
For the measurements presented in this work, preloads of \SIlist{275;450;550}{\degree} have been used, all using the same spring of constant \SI{4.2}{\milli\newton\meter\per\degree}; for this spring, the predicted preload that opposes the pitching moment is \SI{450}{\degree}.
Further considerations on a practical selection of the preload angle are reported in \cref{sec:appendix-preload}.

\subsection{Instrumentation} \label{ssec:instrumentation}
The model-scale tidal turbine has been instrumented to measure the instantaneous angular velocity of the turbine rotor, as well as the instantaneous value of the root-bending moment generated by each individual blade, and the thrust and torque generated by the rotor.
The measurement of rotor angular velocity are obtained by means of an incremental rotary encoder, whose output is both stored and sent to the motor controller to keep the angular velocity of the rotor constant.
The root-bending moment is measured on a per-blade basis, by means of two strain gauges per blade wired in a half-bridge configuration, that are installed on the structural component that links the blades to the turbine rotor.
The thrust and torque generated is instead measured by a force transducer installed between the turbine rotor and the motor shaft; the thrust and torque generated by the turbine are thus not available on a per-blade basis.
The analogue outputs of the strain gauges and the forces transducer are sampled by a National Instruments data acquisition board, while the digital output of the rotary encoder has been sampled by a digital I/O module manufactured by National Instruments.
These quantities are acquired at a frequency of \SI{256}{\hertz}, and each time-series is acquired for a total duration of \SI{300}{\second}.

The incoming velocity observed by the tidal turbine was measured by means of acoustic Doppler velocimetry, with a Vectrino probe that was placed at hub-height and at a streamwise distance of $1.14 D$ upstream of the tidal turbine.
This acquired the free-stream velocity and the water temperature at a frequency of \SI{100}{\hertz} simultaneously with the loads measurements.
The output of this probe has been read before and during each measurement, to ensure that the freestream speed was constant and equal to the mandated $U_\infty$ of \SI{0.8}{\meter\per\second} during all acquisitions.
The same Vectrino probe has been used to measure the temperature of the water in the wave tank $\theta$ during the acquisitions; this measurement is then used to estimate the water density $\rho$ according to \cite{Tanaka2001} and its kinematic viscosity $\nu$ from the estimates of dynamic viscosity $\mu$ according to \cite{Korson1969}.
The relations used are the following:
\begin{linenomath}
    \begin{align}
        \frac{\rho(\theta)}{\rho_{20}} &= \frac{(\theta+a_1)^2(\theta+a_2)}{a_3(\theta+a_4)}, \\
        \log_{10}\left(\frac{\mu(\theta)}{\mu_{20}}\right) &= -\frac{A(\theta-20) + B(\theta-20)^2}{\theta+C},
    \end{align}
\end{linenomath}
where $A = \num{1.1709}$, $B = \SI{1.827e-3}{\per\kelvin}$, $C = \SI{89.93}{\kelvin}$, $\mu_{20} = \SI{1.0020e-3}{\kilo\gram\per\meter\per\second}$ \citep{Korson1969}, $a_1 = \SI{-3.983}{\kelvin}$, $a_2 = \SI{301.797}{\kelvin}$, $a_3 = \SI{522528.9}{\kelvin\squared}$, $a_4 = \SI{69.349}{\kelvin}$, and $\rho_{20} = \SI{999.974}{\kilo\gram\per\meter\cubed}$ \citep{Tanaka2001}.

\subsection{Data reduction} \label{ssec:datareduction}
The data presented in this study consists in the torque $Q$ and thrust $T$ generated by the turbine, as well as the root-bending moment $M_y$ experienced by a single blade, while the turbine operates at a fixed hub-height current velocity $U_\infty = \SI{0.8}{\meter\per\second}$ and at five different tip-speed ratios $\lambda = \numlist{4.5;5;6;7;8}$. The measured $Q$, $T$ and $M_y$ will be presented in the form of coefficients, defined as
\begin{linenomath}
    \begin{align}
        C_Q &=   \frac{Q}{\frac{1}{2} \rho U_\infty^2 \, \pi (D/2)^3}, \\
        C_T &=   \frac{T}{\frac{1}{2} \rho U_\infty^2 \, \pi (D/2)^2}, \\
        C_y &= \frac{M_y}{\frac{1}{2} \rho U_\infty^2 \, \pi (D/2)^3}.    
    \end{align}
\end{linenomath}

Reynolds decomposition is used to separate a time-varying quantity into its time-average value and its zero-mean, fluctuating component.
The time-average will be denoted by an overline and the fluctuating component will be denoted by a prime mark, so that, for instance:
\begin{equation}
    C_Q(t) = \overline{C_Q} + C_Q'(t), 
\end{equation}
where the time-variation of $C_Q(t)$ has been removed in $\overline{C_Q}$ by virtue of time-averaging, and $\overline{C_Q'} = 0$.

In \cref{tab:uncertainties} we include the half-width of the \SI{95}{\percent} confidence intervals around the mean of all quantities presented in this paper.
These are moreover subdivided in those that are directly measured (the forces and torques acting on the turbine, the turbine angular velocity, the temperature of the water and the free current velocity) and the derived ones (water density and viscosity, tip-speed ratio, force and torque coefficients).
The detailed derivation of each value of uncertainty is reported in \cref{sec:appendix-uncert}.

\begin{table}[ht!]
    \centering
    \caption{Width of the \SI{95}{\percent} confidence intervals on the means of all measurements.}
    \label{tab:uncertainties}
    \vspace{6pt}
    \begin{tabular}{ll} \toprule
        Quantity                              & \SI{95}{\percent} confidence interval        \\ \midrule
        Temperature $\theta$                  & $\pm$  \SI{2e-1}{\kelvin}                    \\
        Free-current velocity $U_\infty$      & $\pm$  \SI{1e-2}{\meter\per\second}          \\
        Turbine angular velocity $\omega$     & $\pm$  \SI{5e-1}{\revolution\per\minute}     \\
        Torque $Q$                            & $\pm$  \SI{3e-1}{\newton\meter}              \\
        Thrust $T$                            & $\pm$  \SI{4e+0}{\newton}                    \\
        Root-bending moment $M_y$             & $\pm$  \SI{6e-1}{\newton\meter}              \\ \midrule
        Density $\rho$                        & $\pm$  \SI{4e-2}{\kilo\gram\per\meter\cubed} \\
        Kinematic viscosity $\nu$             & $\pm$  \SI{5e-6}{\meter\squared\per\second}  \\
		Tip-speed ratio $\lambda$             & $\pm$ \num{8e-2}                             \\
		Torque coefficient $C_Q$              & $\pm$ \num{2e-3}                             \\
		Thrust coefficient $C_T$              & $\pm$ \num{3e-2}                             \\
		Root-bending moment coefficient $C_y$ & $\pm$ \num{6e-3}                             \\ \bottomrule
    \end{tabular}
\end{table}

\section{Results}\label{sec:results}
\subsection{Rigid blade reference} \label{ssec:res_rigid}
The turbine performance, characterised by the thrust and torque generated by the rotor and the root-bending moment generated by the individual blades, has been measured for tip-speed ratios of \numlist{4.5;5;6;7;8}, where $\lambda = \num{4.5}$ is the value for which the turbine generates the maximum torque.
This section presents measurements of these quantities for a turbine rotor equipped with rigid blades, to provide a benchmark with which to compare the results obtained for passively pitching blades.
The effectiveness of the passive pitch mechanism will be assessed with respect to two aspects, namely the load mitigation performance and the average load generated, which will be presented in \cref{ssec:res_passive}.

\begin{figure}[ht!]
    \centering
    \includegraphics{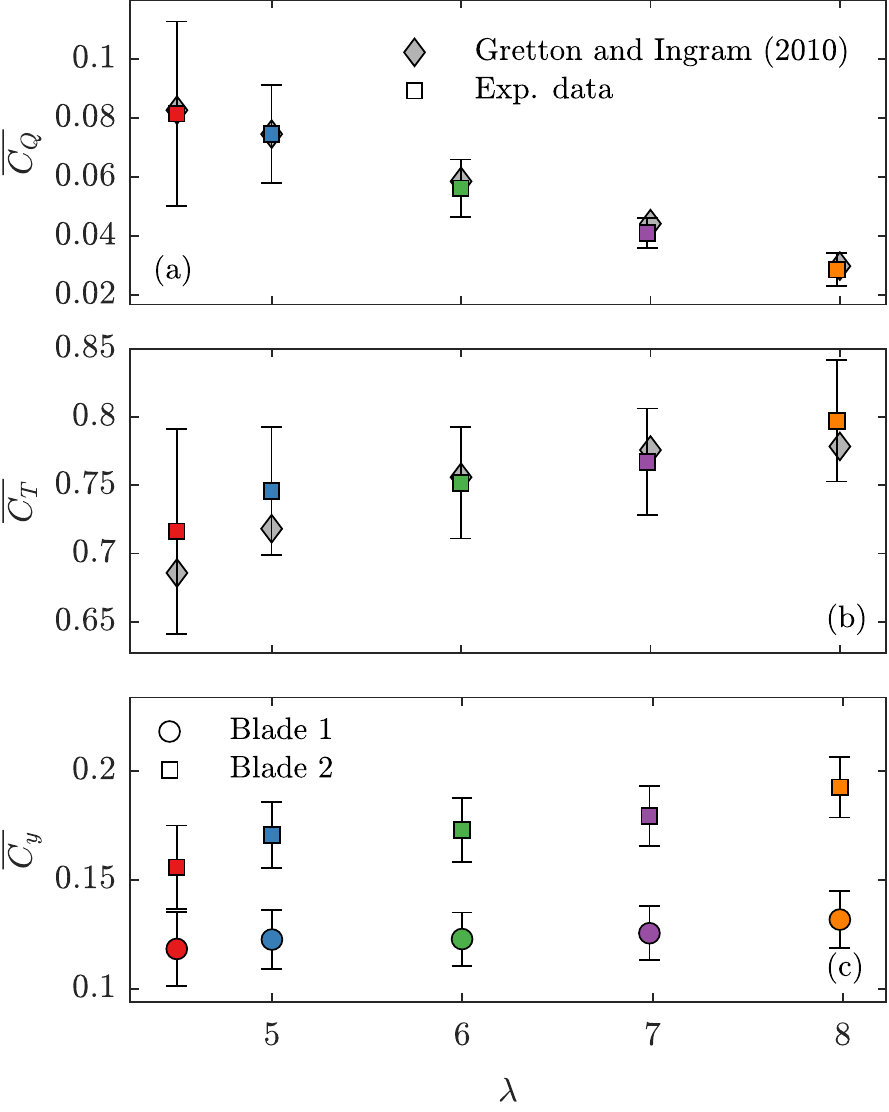}
    \caption{From top to bottom: mean values of torque $\overline{Q}$, thrust $\overline{T}$ and root-bending moment $\overline{M_y}$ for the turbine equipped with rigid blades. Squared markers are filled with different colours for each value of the tip-speed ratio $\lambda$. The load fluctuations are represented by one standard deviation half-width error bars. Experimental data is compared to the simulations of \cite{Gretton2010} for the torque and thrust coefficients (grey diamonds).}
    \label{fig:rigidblade-meanvalues}
\end{figure}
\Cref{fig:rigidblade-meanvalues} shows the mean rotor torque coefficient $\overline{C_Q}$ and thrust coefficient $\overline{C_T}$, and the rigid blade bending moment coefficient $\overline{C_y}$ for all the tested tip speed ratios $\lambda$ at the nominal flow speed $U_\infty = \SI{0.8}{\metre\per\second}$.
The values of torque and thrust measured experimentally are also compared to those estimated via an in-house blade element-momentum code reported by \citet{Gretton2010} for the same turbine geometry: these results show good agreement, especially in the generated torque.
The scatter dots show the average values while the error bars have a half-width equal to one standard deviation of each load time-history. 
The thrust and root-bending moment show a similar trend, with the average values increasing at higher $\lambda$.

It can be noted that the amplitude of the torque fluctuations is somewhat proportional to the mean torque value, being higher at $\lambda = 4.5$ and almost negligible at $\lambda$ of \numlist{7;8}.
On the other hand, it can be seen that both the turbine torque and the root-bending moment generated by the blades show fluctuations with magnitudes that are approximately constant for all values of $\lambda$. 
As for the values of root-bending moment generated, we report the values of blade \num{1} and blade \num{2} separately as the installation issue described in \cref{sec:method} resulted in a different angle of geometrical pitch for this blade.
Moreover, the measurements of root-bending moment for blade \num{3} are not available due to a fault in the root-bending flexures for this blade.

\begin{figure}[ht!]
    \centering
    \includegraphics{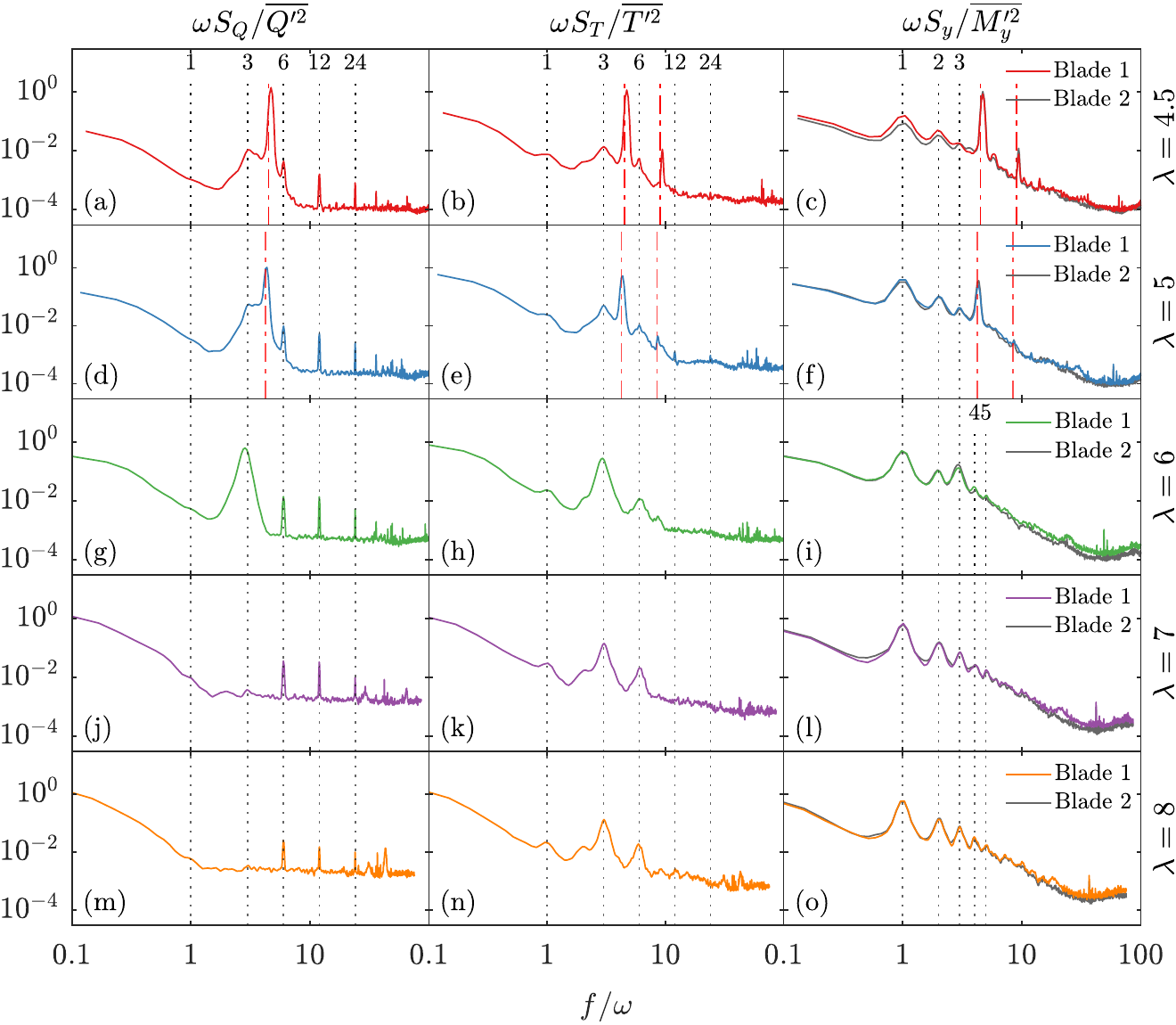}
    \caption{From left to right: spectra of torque $S_Q$, thrust $S_T$, and root-bending moment $S_y$ for the five operating conditions investigated (top to bottom), plotted \textit{versus} frequency normalised by the turbine angular velocity $\omega$. The dotted vertical lines denote the frequency of relevant harmonics, while the red dashed lines indicate harmonics that are invariant to the angular velocity.}
    \label{fig:rigidblade-spectra-all}
\end{figure}
\Cref{fig:rigidblade-spectra-all} shows the spectra of the fluctuations in all measured loads (torque, thrust and root-bending moment) for the turbine equipped with rigid blades, as a function of the frequency axis $f$ of the Fourier domain.
Some harmonics, namely the one at the turbine angular velocity, and the one at the blade-passing frequency and its higher-order harmonics, are highlighted by vertical dashed lines.
Moreover, two harmonic components of the loading at two fixed frequencies are highlighted, for values of $\lambda < 5$, by red dash-dotted lines: these are located at the dimensional frequencies of \SIlist{4.5;9}{\hertz} for all test cases.

The composition of the fluctuations in the frequency domain is markedly different between quantities and tip-speed ratios.
Starting from the torque spectra and ignoring the harmonics whose frequencies are invariant with $\omega$, denoted by the red lines in the figure, one can note how the main contributor to the fluctuations in torque is a relatively narrow-band component at the blade-passing frequency $3\omega$, due to each blade rotating in the sheared flow moving from high-speed regions above the hub to low-speed regions below the hub, where the flow velocity is further affected by the presence of the turbine tower which deviates the upstream flow.
It is interesting to note how the fluctuations at high tip-speed ratios (\cref{fig:rigidblade-spectra-all}(j,m)) are instead dominated by stochastic components at frequencies lower than $\omega$, as the torque generated by the blades decreases with increasing $\lambda$.
Three additional very-narrow-band components are observed for all values of $\lambda$ for $f/\omega = \numlist{6;12;24}$: these have been previously found to be caused by motor cogging due to the 12-pole brushless DC machine used during these experiments \citep{Noble2020}, and thus are of no interest to the analysis here presented.

The thrust fluctuations are similarly dominated by narrow-band fluctuations at $3\omega$ and their higher-order harmonics: in this case, the magnitude of these components increases with $\lambda$ as the mean torque generated by the turbine is also increasing.
An additional component at $\omega$ is present at all values of $\lambda$, as it is likely due to either a drivetrain misalignment or load imbalance between blades: as blade 2 has effectively operated at a higher pitch angle, it has generated a lower thrust and thus an unequal loading around the rotor.

The root-bending moment fluctuations are instead dominated by a component at $\omega$: this is expected, as the root-bending moment is measured for each blade individually, as opposed to the measured torque and thrust which are instead the sum of the individual blades loading.
This peak is present, along with its first two harmonics, for all values of tip-speed ratio, while harmonics up to the fourth, at $5\omega$, are visible at high tip-speed ratios.
It is interesting to note that, despite the difference in pitch angle and therefore on the mean root-bending moment generated by the blades, the measured spectra are markedly similar and seemingly unaffected by the actual pitch angle of the blade.

\begin{figure}[ht!]
    \centering
    \includegraphics{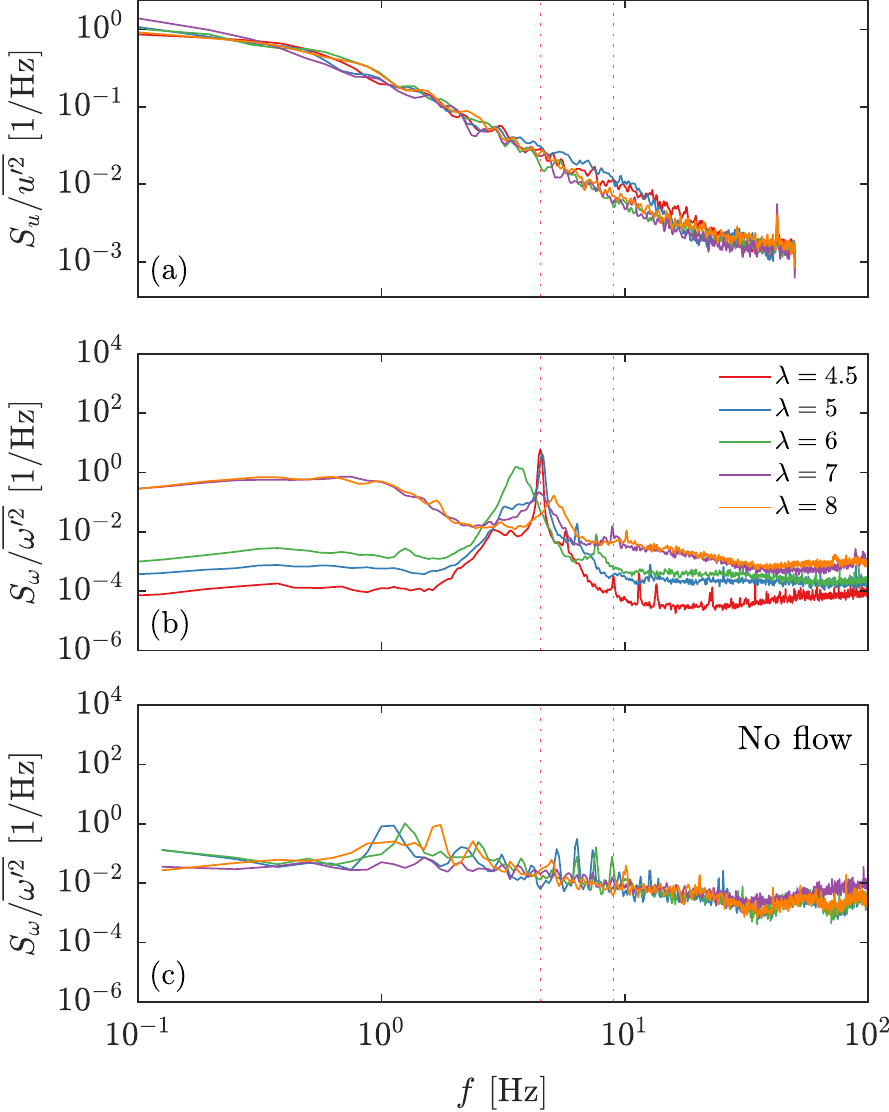}
    \caption{From top to bottom: spectra of incoming velocity $S_u$ (\textit{a}) and spectra of the angular velocity fluctuations $S_\omega$ with incoming flow on the turbine (\textit{b}) and without incoming flow on the turbine (\textit{c}): colours between (\textit{b}) and (\textit{c}) match the mean angular velocity between test cases. The red dotted lines are at the same frequency as the ones in \cref{fig:rigidblade-spectra-all}. Note both axes in dimensional units.}
    \label{fig:rigidblade-instability}
\end{figure}
The contributions to the spectra marked with the red dash-dotted lines in \cref{fig:rigidblade-spectra-all} can be explained by analysing the spectra of the turbine angular velocity fluctuations, reported in \cref{fig:rigidblade-instability}(b): along with these, the figure also reports the spectra of the incoming velocity fluctuations and the spectra of angular velocity fluctuations when the turbine is not subject to an incoming velocity field; the two values of frequency for which the unknown contributions to the spectra were observed are reported with their dimensional values of \SIlist{4.5;9}{\hertz}.
As the peaks appear in the spectra of angular velocity at the same dimensional frequency for different values of the turbine tip-speed ratio, namely \numlist{4.5;5}, one can conclude that these are not representative of fluid-structure interaction phenomena: this is further highlighted by the absence of components at these frequencies in the spectrum of incoming velocity $S_u$ plotted in \cref{fig:rigidblade-instability}(a).
A peak at a dimensional frequency of \SI{4.5}{\hertz} is also observed in the angular velocity spectrum for a $\lambda = 7$; however, in this case, this frequency corresponds to $3\omega$ and it is thus caused by the rotation of the turbine in a sheared inflow.
Since the peaks are not visible when the turbine rotates in the absence of current (compare with the \textit{No flow} spectra of \cref{fig:rigidblade-instability}(c)), the phenomenon that gives rise to these components of the spectrum is likely an instability in the turbine angular velocity that happens for specific combinations of $\omega$ and $U_\infty$, as previously noted by \cite{Payne2017}.

\subsection{Passively pitching blade} \label{ssec:res_passive}
In \cref{sssec:effects_preload}, we present the time-averaged values of the torque, thrust, and root-bending moment and their standard deviation as a measure of their unsteadiness for the turbine equipped with two rigid and one passively pitching blade.
This data will show that the mean value of torque, and thus power, generated by the turbine is conserved or only weakly reduced by the passive pitch mechanism.
The effects of the passive pitching system on the unsteadiness of the loads is instead reported in the following \cref{sssec:loadmitigation}: this will show that the passive control system is well suited to mitigate the unsteady components of the loads.
The time-averaged values of the turbine performance coefficients are shown for three distinct values of preload, namely $\beta_\mathrm{pre} = \SIlist{275;450}{\degree}$ for blade 1 and $\beta_\mathrm{pre} = \SI{550}{\degree}$ for blade 2.
The turbine is operated, for ease of comparison with the rigid blade case, at the same values of tip-speed ratio $\lambda$ as the previous section.

\subsubsection{Effects of the preload angle} \label{sssec:effects_preload}
\begin{figure}[ht!]
    \centering
    \includegraphics{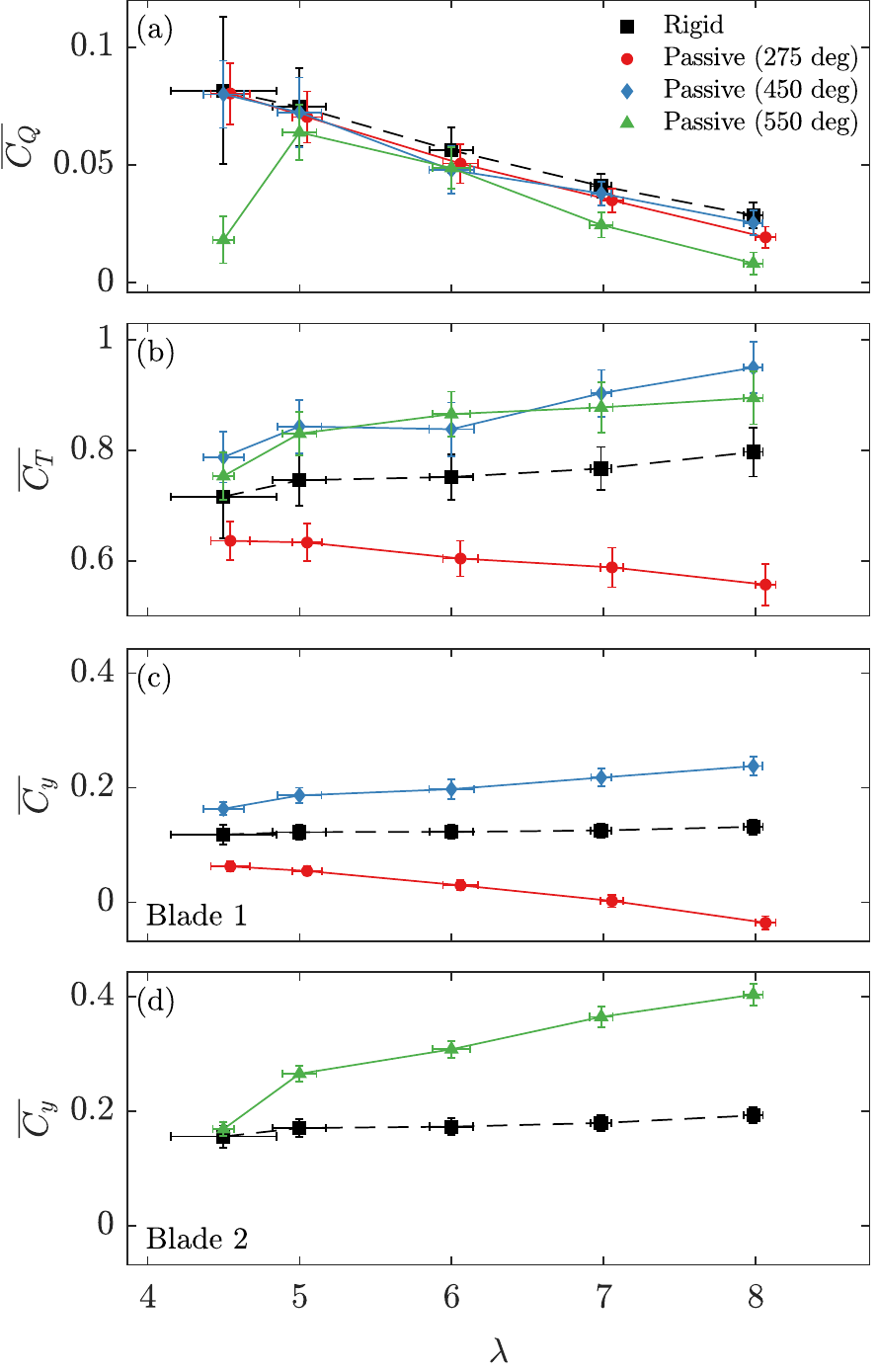}
    \caption{Mean values of torque $Q$, thrust $T$ and root-bending moment $M_y$ attained with the rotor fitted with one passively pitching blade (\textit{coloured lines}) and with only rigid blades (\textit{black line}). The error bars half-width is equal to one standard deviation of each measured quantity.}
    \label{fig:passivepitch-meanvalues}
\end{figure}
\Cref{fig:passivepitch-meanvalues} reports the mean values of thrust, torque and root-bending moment attained by the turbine when fitted with one passively pitching blade.
It is worth recalling that the optimal value for $\beta_\mathrm{pre}$ of \SI{450}{\degree} was computed under the assumption of $\lambda = 7$ and negligible friction.
Data acquired shows that the torque generated by the turbine, and thus ultimately the power harvested, are minimally sensitive to the spring preload angle $\beta_\mathrm{pre}$, as all tested preloads result in comparable values of $C_Q$ for all tip-speed ratios tested, with the only exception being the high-preload test case of $\beta_\mathrm{pre} = \SI{550}{\degree}$ at $\lambda = 4.5$.
Moreover, one can see that there is an excellent agreement between the torque generated at the on-design conditions of $\lambda = 7$ and $\beta_\mathrm{pre} = \SI{450}{\degree}$ and those of the turbine equipped with rigid blades, confirming the goodness of the reduced-order approach to predict the actual preload.
In addition, it can be seen that the torque generated by the turbine equipped with one pitching blade is comparable to that generated by a fully rigid rotor.
The most immediate consequence of this is that the time-averaged power generated by a turbine equipped with passively pitching blades is comparable to that of a canonical tidal turbine, as predicted by the reduced-order model.

The values of $C_T$ and $C_y$ are instead seen to be more affected by changes in the preload angle than the respective values of $C_Q$.
In fact, a clear distinction can be seen between the test case at $\beta_\mathrm{pre} = \SI{275}{\degree}$, for which the torque and the root-bending moment are less than those of a rigid-bladed rotor, and the higher-preload test cases.
In addition to this, it must be noted that these results highlight a potential advantage of the passive-pitch system that had not been evident from the reduced-order approach: for values of preload much lower than design, the torque and power generated by the turbine are not significantly affected, while the thrust and root-bending moment are reduced.
This suggests that the time-averaged structural loads on the turbine structure and on the blade can be reduced with minimal effect on the turbine power production by allowing the blades to passively pitch.

The effects of the preload angle on the mean turbine performance can be understood as this effectively sets the time-average value of the blade pitch: a too high value of $\beta_\mathrm{pre}$ induces a moment larger than the hydrodynamic pitching moment generated by the current is so that the blade pitches with its leading edge downstream, thus changing its angle of attack and increasing the hydrodynamic moment generated by the current until an equilibrium is found; as the mean angle of attack of the blades increases, so does the load generated and thus so does the turbine thrust and the root-bending moment, which is visualised as a larger-than-baseline values of both $T$ and $M_y$.
To the limit, this can result in the blade pitching to an angle of attack larger than that of stall, especially for low tip-speed ratios $\lambda$: this can result in separated flow around the blade, which in turn causes large unsteadiness in the blade loads and low time-average values of torque and thrust; while the instantaneous pitching angle of the blade has not been measured, it can be reasonably assumed this is the reason behind the low values of both $Q$ and $M_y$ at $\lambda = 4.5$ for the $\beta_\mathrm{pre} = \SI{550}{\degree}$ test case.
Vice versa, smaller values of the preload angle result in a spring-back moment smaller than the hydrodynamic one, with the blades pitching their leading edge upstream and decreasing their angle of attack, resulting in lower values of $T$ and $M_y$ than the reference.
These results also show that the correct value is lower than the \SI{450}{\degree} predicted by the low-order code.
As outlined in \cref{eq:ss_pitching}, the value of $\beta_\mathrm{pre}$ is affected by the spring stiffness, the hydrodynamic pitching moment around the blades and the static friction that opposes the pitching moment.
It will therefore be possible to achieve better estimates of the necessary preload angle by including an estimate of the friction drag in the reduced-order code.

To further interpret these results, let us consider that the design preload angle was \SI{450}{\degree}, and it was computed for an $U_\infty = \SI{0.8}{\meter\per\second}$ and a $\lambda = 7$.
Experiments showed that preload angles of \SIlist{450;550}{\degree} are too high, and \SI{275}{\degree} is too low.
For higher preloads, the blade experiences higher angles of attack along its entire span compared to design conditions, whereas for the lower preload the incidence is lower.
When operating at higher tip speed ratio, $M_y$ changes according to two competing effects, namely dynamic pressure increase and reduction of the angle of attack, with the former intensifying the average $M_y$, and the latter reducing it.
Both the dynamic pressure and the incidence change equally on each blade, regardless of the spring preload; however, their effect on the $M_y$ is different.
Considering the lift developed by a blade section at a given angle of attack $\alpha$, and approximating $C_l \approx 2\pi\alpha$, one has:
\begin{equation}
    L = \frac{1}{2} \rho U^2 c \, 2 \pi \alpha.
\end{equation}
For increasing values of $\lambda$, the inflow speed increases by $\Delta U$ and the incidence decreases by $\Delta \alpha$, which are independent of the preload and hence equal for each blade.
In first approximation, a change in lift can be caused by either a change in incoming velocity $\Delta U$ or a change in angle of attack $\Delta \alpha$ as 

\begin{equation}
    \Delta L = \frac{\partial L}{\partial U} \Delta U +
               \frac{\partial L}{\partial \alpha} \Delta \alpha, 
\end{equation}
where
\begin{linenomath}
    \begin{align}
        \frac{\partial L}{\partial U} &= \rho U c \, 2 \pi \, \alpha, \\
        \frac{\partial L}{\partial \alpha} &= \frac{1}{2} \rho U^2 c \, 2 \pi .
    \end{align}
\end{linenomath}

For any preload, a change in the angle of attack causes the same reduction in lift, while an increase in the dynamic pressure will have larger effects the larger the initial load on the blade is. 
It is therefore possible that, for preloads of \SIlist{450;550}{\degree} which have resulted in high values of $\alpha$ along the blade, the dominant component of $\Delta L$ is the one accounting for the change in dynamic pressure, in turn causing the root-bending moment to increase with $\lambda$.
Vice versa, the lower $\alpha$ attained for $\beta_\mathrm{pre} = \SI{275}{\degree}$ might have resulted in the $\partial L/\partial \alpha$ component to be prevalent, in turn resulting in a decreasing root-bending moment with $\lambda$.

\subsubsection{Load mitigation} \label{sssec:loadmitigation}
\begin{figure}[ht!]
    \centering
    \includegraphics{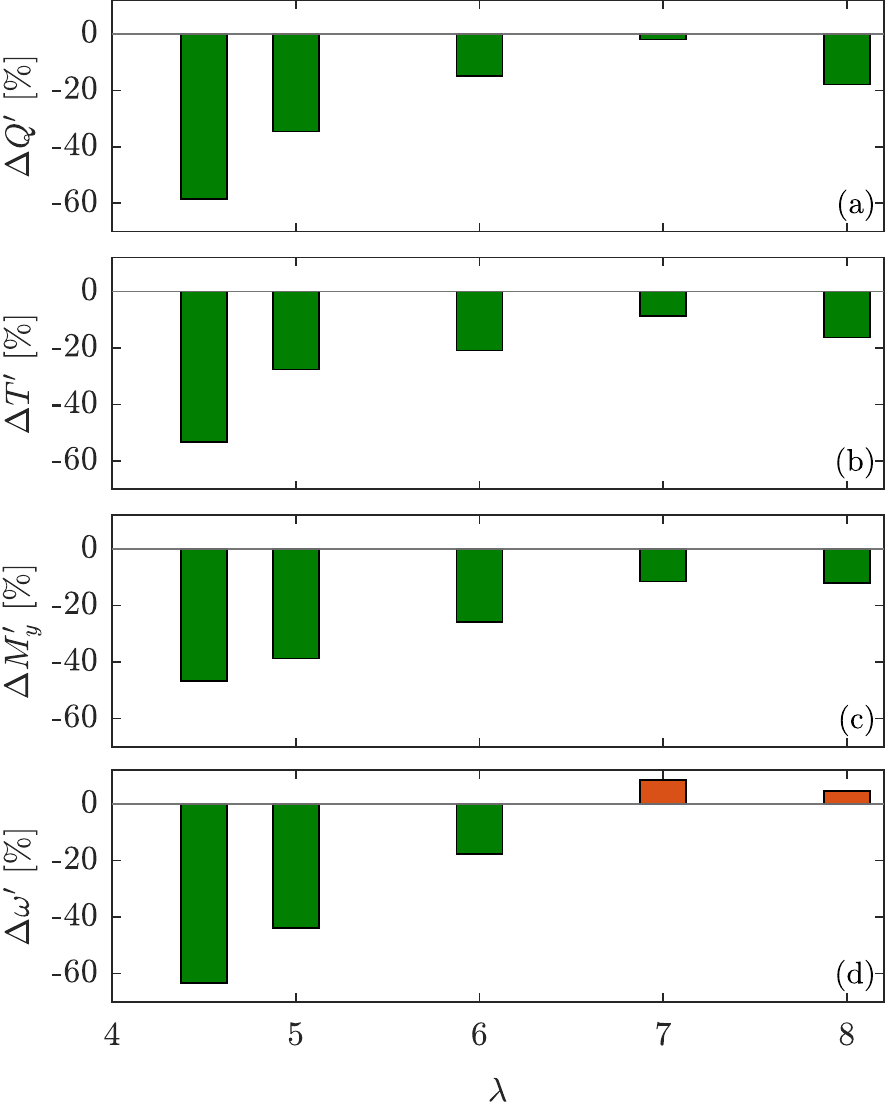}
    \caption{Mitigation of the torque, thrust and root-bending moment fluctuations measured as the relative difference between standard deviations of these quantities with respect to the fully-rigid rotor. Blade spring preloaded by \SI{275}{\degree}.}
    \label{fig:passivepitch-std}
\end{figure}
\Cref{fig:passivepitch-std} reports the relative difference of the fluctuations standard deviations between rigid and passively pitching rotor blades, obtained for a spring preload of \SI{275}{\degree}.
The relative difference for the torque generated by the rotor is defined as
\begin{equation}
    \Delta Q' = \frac{\sqrt{\overline{Q'^2_\mathrm{pitch}}} - \sqrt{\overline{Q'^2_\mathrm{rigid}}}}{\sqrt{\overline{Q'^2_\mathrm{rigid}}}}
\end{equation}
where $\sqrt{\overline{Q'^2_\mathrm{rigid}}}$ is the standard deviation of the torque timeseries acquired with a rigid rotor, and $\sqrt{\overline{Q'^2_\mathrm{pitch}}}$ is that of the passively pitching blade; the other quantities in \cref{fig:passivepitch-std} are defined accordingly.
In all cases considered, all the loads have consistently been reduced.
Peak performance can be observed at $\lambda = 4.5$, where fluctuations of thrust are mitigated by \SI{55}{\percent}, root-bending moment fluctuations by \SI{45}{\percent}, and torque fluctuations by \SI{60}{\percent}. The worst performance is observed at $\lambda = 7$, where fluctuations of thrust, root-bending moment, and torque are still reduced by \SI{7}{\percent}, \SI{14}{\percent}, and \SI{2}{\percent} respectively.
Note that there is no value of tip-speed ratio for which the passive pitching system causes an increase in the standard deviation of the loads experienced by the turbine, confirming the goodness of this technology is not limited to the design conditions but it is more broadly applicable to a wide range of operating conditions.
Data for these plots is not reported for the preload angles of \mbox{\SIlist{450;550}{\degree}} as these resulted in a non-negligible bend in the spring axes, which in turn resulted in the spring inner diameter to come in contact with the pitching shaft.
This has added friction to the system in a quantity that is not trivial to estimate, and it has consequently affected the performance of the passive pitch system.

Let us note that the passive pitch performance is consistent with the changes of the turbine speed fluctuations shown in \cref{fig:passivepitch-std}, as load fluctuations are due to oscillations of the blade incidence but also to fluctuations of the dynamic pressure, which is determined mainly by the turbine speed.
In particular, at low tip speed ratio ($\lambda = \numrange{4.5}{6}$) the oscillations of the loads and of the turbine speed are both alleviated.
Since the passive pitch system mitigates the loads by changing the blade pitch angle, it affects the blade incidence directly and the turbine speed indirectly, which instead is determined by a more complex interaction between the rotor, the drivetrain, and the control system.
It is thus not straightforward to determine if the load mitigation was due to the change of the turbine speed or to the passive pitch system.
On the other hand, at high tip speed ratio ($\lambda = \numlist{7;8}$), the loads are alleviated despite the increase in amplitude of the turbine speed fluctuations, proving that the load alleviation is not necessarily a direct result of a more steady turbine speed.
This however hints at a different mechanism underlying the mitigation of load fluctuations for low- and high-$\lambda$: to analyse this more in detail it is useful to observe the behaviour of the turbine for these two different conditions separately.

\begin{figure}[ht!]
    \centering
    \includegraphics{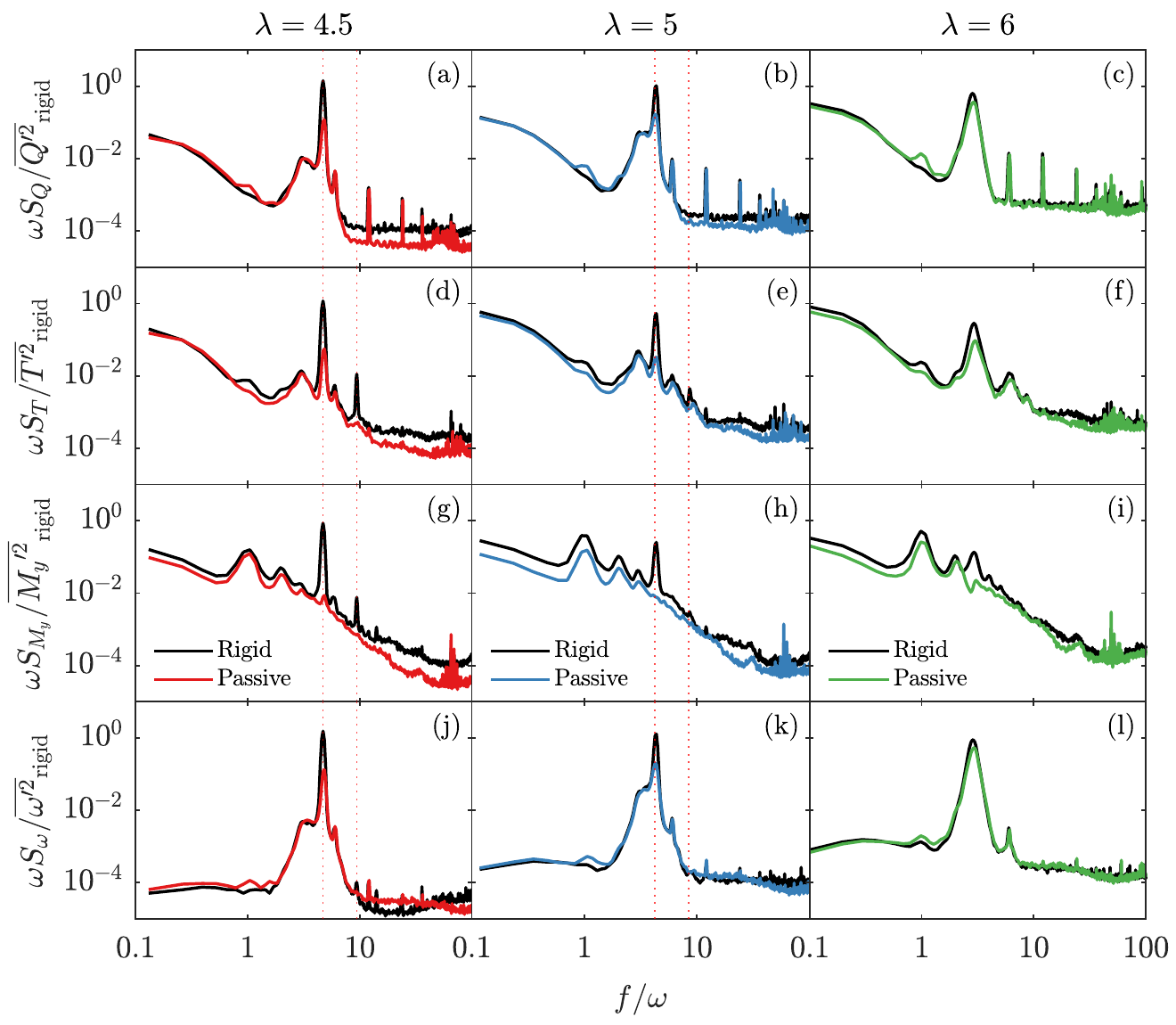}
    \caption{Spectra of torque, thrust, root-bending moment and angular velocity (\textit{top to bottom}) for tip-speed ratios $\lambda = \numlist{4.5;5;6}$ (\textit{left to right}), for a turbine equipped with rigid blades (\textit{black lines}) and passively pitching blades (\textit{coloured lines}), along with $\omega$-independent loads (\textit{dotted red lines}). Each spectrum is normalised with the variance of the quantity measured with a rigid-blade rotor.}
    \label{fig:passivepitch-spectra-lowtsr}
\end{figure}
\Cref{fig:passivepitch-spectra-lowtsr} reports the spectra of the torque, thrust, root-bending moment and angular velocity obtained with both the passive pitch rotor and the one equipped with rigid blades, for the tip-speed ratios under \num{6}.
These are the values of $\lambda$ for which $\Delta \omega'$ is negative, signalling that the load fluctuations mitigation is concurrent with a reduced unsteadiness of the turbine angular velocity.
Note that, to warrant a meaningful comparison between the spectra of the same quantity obtained with a rigid and a passively pitching rotor, all spectra are normalised by the variance of the relevant quantity obtained at the given tip-speed ratio with a rigid-blade rotor: this means that the plotted spectra obtained with passively pitching blades do not integrate to one over the frequency axis, which is representative of a lower total magnitude of the fluctuations.

Where present, the intensity of the highest peak at \SI{4.5}{\hertz}, which is due to turbine-drivetrain instability as shown in \cref{ssec:res_rigid}, is reduced by at least one order of magnitude for all loads.
In the case of the root-bending moment (\cref{fig:passivepitch-spectra-lowtsr}(g,h,i)), the peak is reduced by two orders of magnitude, to the point it can no longer be discerned from the surrounding components due to the current turbulence.
However, the spectrum of the turbine speed shows a comparable mitigation of the same peak, suggesting that the load alleviation is due to a more stable speed rather than the passive pitch system.
The turbine speed presents also a higher peak at $f/\omega = 1$, which can also be noted in the torque spectrum, caused by the rotor imbalance due to the passively pitching blade that experiences different loads compared to the other two blades.
Beside the above observations, the differences between the spectra of the passive pitch case and those of the rigid case are quite small, which makes it difficult to judge the effectiveness of the passive pitch system across the entire spectrum.

The effects of the passive pitching system on the loads experienced on the wind turbine can be more thoroughly analysed in the spectral domain by means of a transfer function: this is defined for the torque as
\begin{equation}
    G_Q(f) = \frac{S_{Q_\mathrm{pitch}}(f)}{S_{Q_\mathrm{rigid}}(f)}
\end{equation}
where $S_{Q_\mathrm{rigid}}$ is the spectrum of the torque fluctuations for a rigid-bladed rotor, and $S_{Q_\mathrm{pitch}}$ refers to that of a rotor equipped with one passive blade, and accordingly for all other quantities.
The value of the transfer function is thus lower than unity at frequencies for which the passive pitching mechanism lowers the magnitude of the load fluctuations and greater than one otherwise.

\begin{figure}[ht!]
    \centering
    \includegraphics{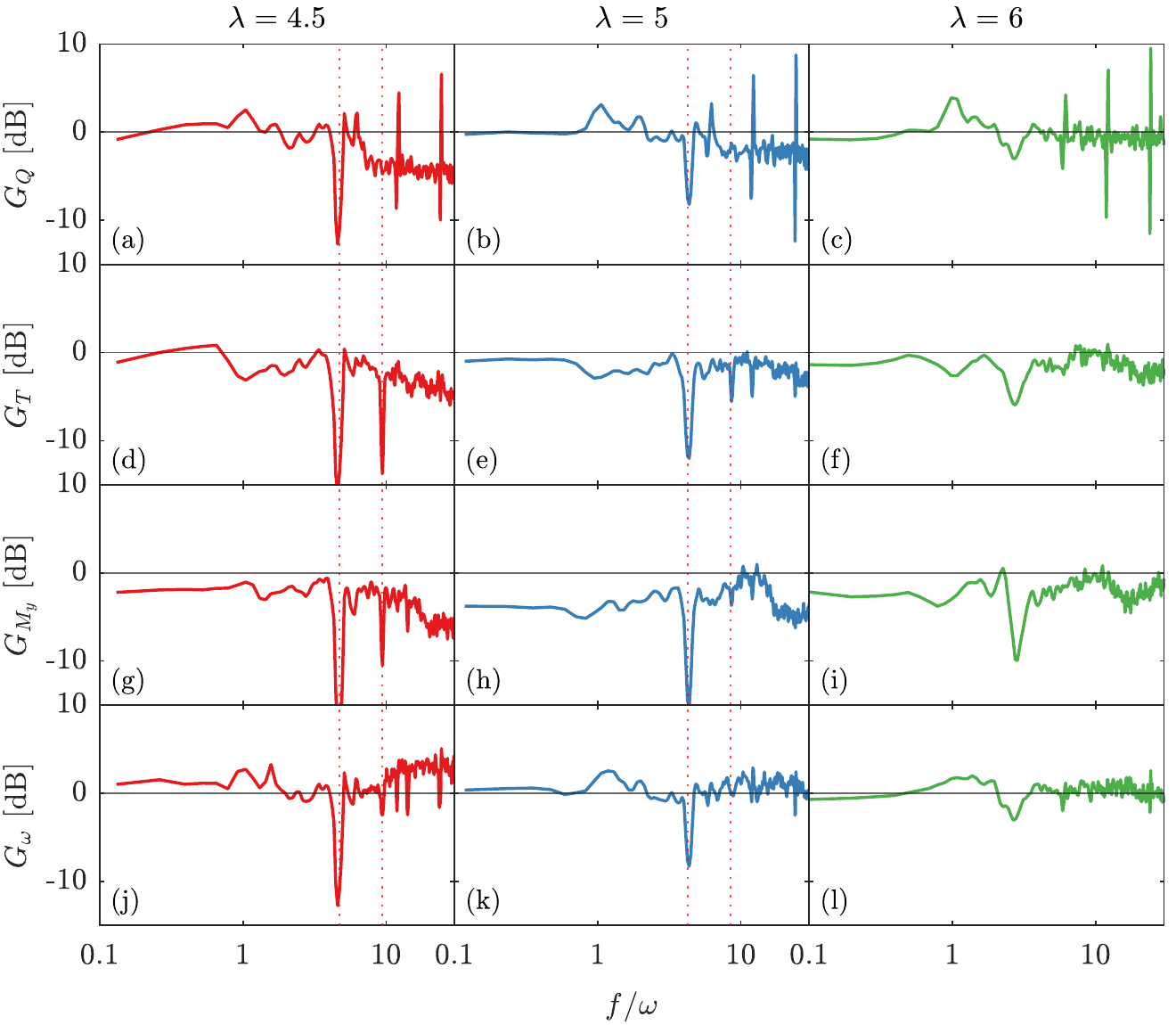}
    \caption{Transfer functions of torque, thrust, root-bending moment and angular velocity (\textit{top to bottom}) acquired at tip-speed ratios of \numlist{4.5;5;6} (\textit{left to right}). Transfer functions relative to the spectra pairs presented in \cref{fig:passivepitch-spectra-lowtsr}.}
    \label{fig:passivepitch-trf-lowtsr}
\end{figure}
The transfer functions for the torque, thrust, and root-bending moment loads as well as the turbine angular velocity are plotted for the low-$\lambda$ test cases in \cref{fig:passivepitch-trf-lowtsr}.
The most immediate observation that can be drawn is that of the reduction in the loads that are generated by drivetrain instability at the dimensional frequencies of \SIlist{4.5;9}{\hertz}, which are denoted by the red dotted vertical lines.
However, apart from these contributions, it can be seen that the passive pitch mechanism is effective in reducing the root-bending moment of the turbine blade along the whole frequency spectrum, regardless of the tip-speed ratio, as the transfer function $G_{M_y}$ is always lower than \SI{0}{\deci\bel}; similar conclusions are drawn by observing the thrust transfer function $G_T$.
The effects are larger for the root-bending moment, since it considers only the passively pitching blade and is therefore more representative of the effectiveness of the passive pitch system.

The mitigation of the load fluctuations is thus due in part to the reduction of the turbine speed fluctuations, and in part to the passive pitch system.
Most importantly, the data shows that the root-bending moment fluctuations at $\omega$ and higher-order harmonics is mitigated by the passive pitch system, despite greater speed oscillations.
Since the system was designed to alleviate load fluctuations at those frequencies (i.e. oscillations due to shear flow), these results confirm the load-mitigating capabilities of the passive pitch system.

\begin{figure}[ht!]
    \centering
    \includegraphics{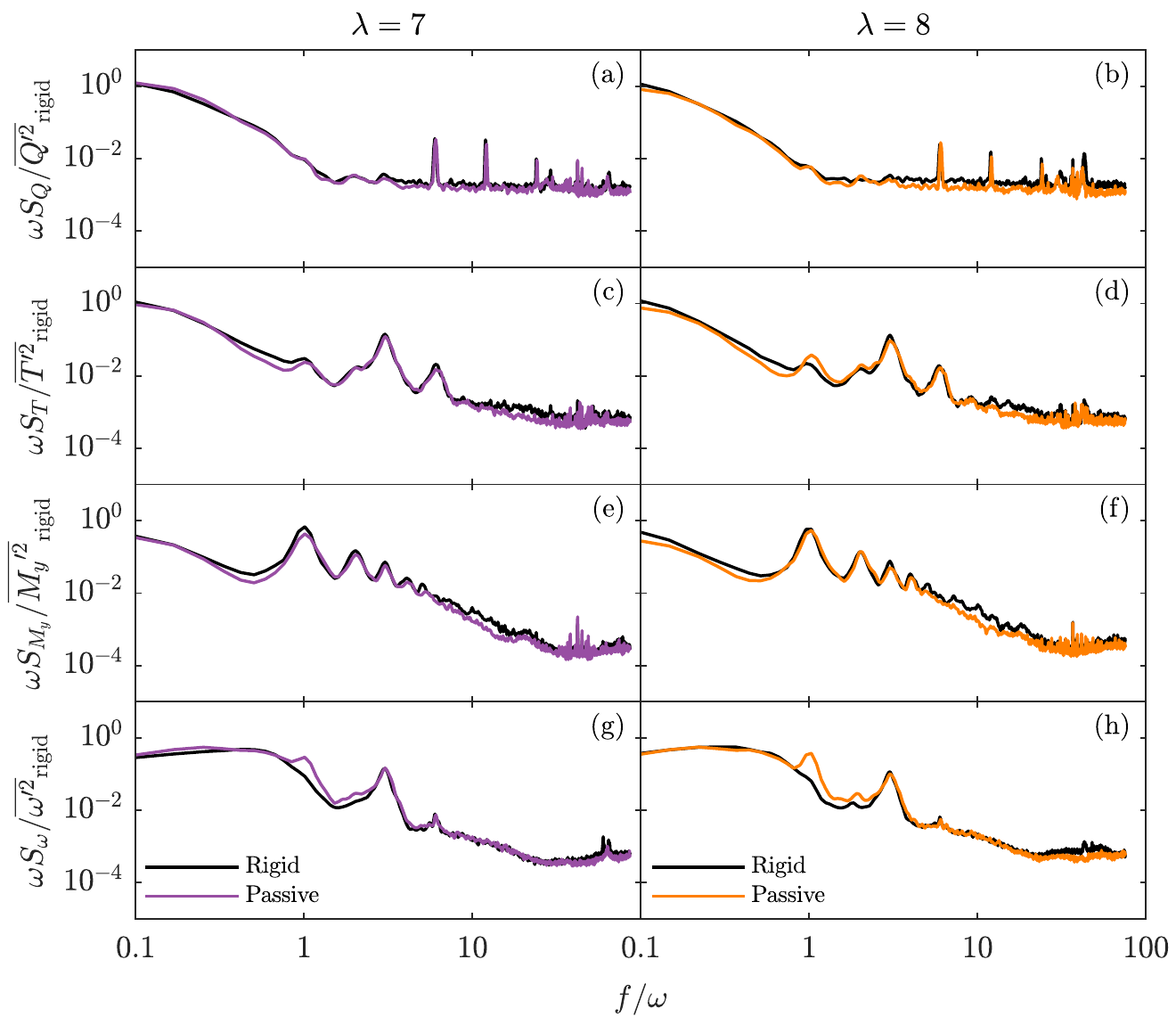}
    \caption{Spectra of torque, thrust, root-bending moment and angular velocity (\textit{top to bottom}) for tip-speed ratios $\lambda = \numlist{7;8}$ (\textit{left to right}), for a turbine equipped with rigid blades (\textit{black lines}) and passively pitching blades (\textit{coloured lines}).}
    \label{fig:passivepitch-spectra-hightsr}
\end{figure}
\Cref{fig:passivepitch-spectra-hightsr} shows the loads and turbine speed spectra for a turbine operating with one passively pitching blade, at $\lambda = \numlist{7;8}$ respectively.
For these test cases, no peak at \SI{4.5}{\hertz} is found.
As opposed to lower tip speed ratios, there is no obvious peak reduction, and the turbine speed is dominated by low frequency fluctuations.
In particular, the turbine speed spectra shows a more intense peak at $\omega$ due to load imbalance, as only one blade is pitching, which can explain the positive value of $\Delta \omega'$ for these test cases.
Overall, the differences between the spectra for the passive pitch case and those for the rigid case are small.

\begin{figure}[ht!]
    \centering
    \includegraphics{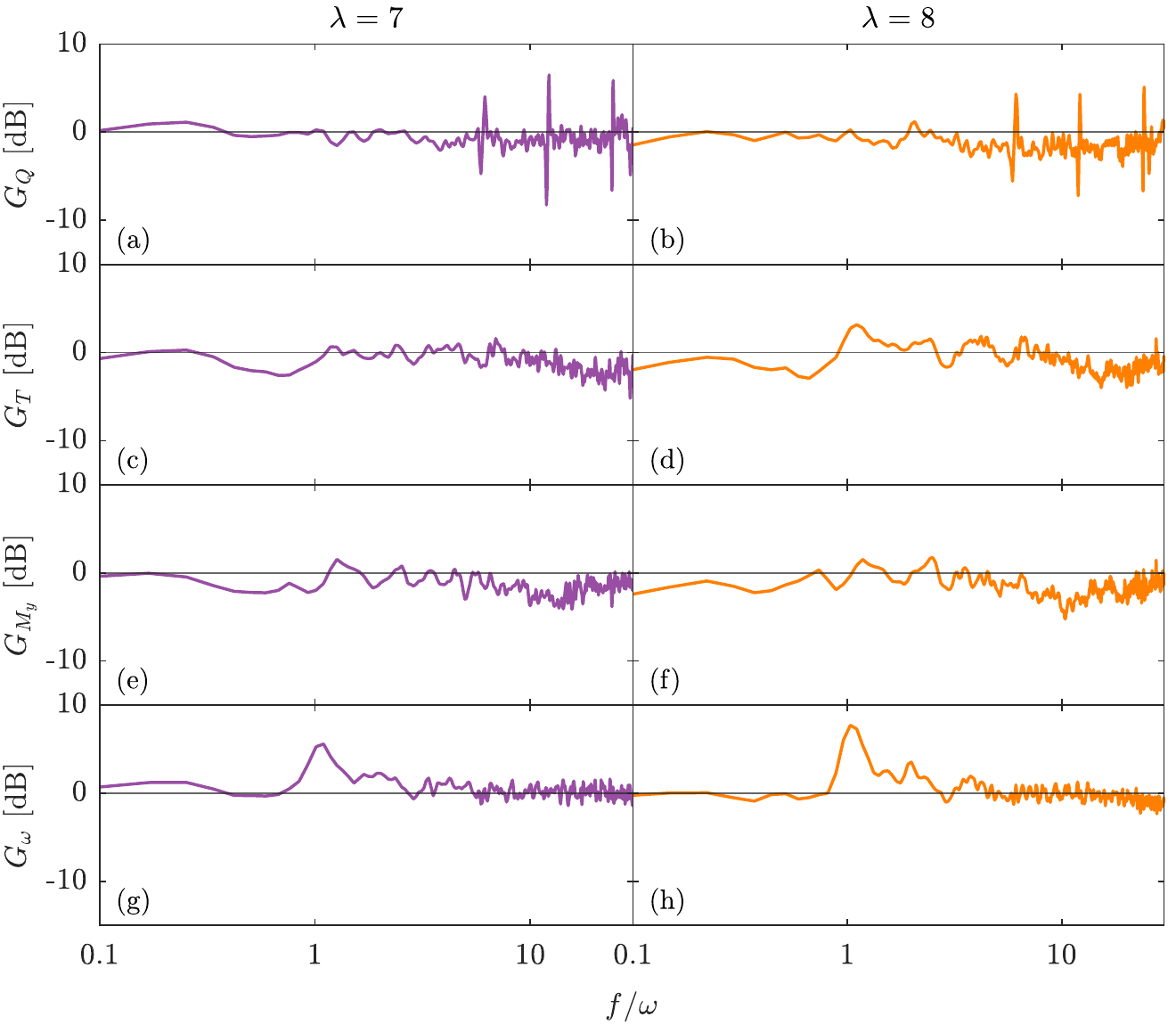}
    \caption{Transfer functions of torque, thrust, root-bending moment and angular velocity (\textit{top to bottom}) acquired at tip-speed ratios of \numlist{7;8} (\textit{left to right}). Transfer functions relative to the spectra pairs presented in \cref{fig:passivepitch-spectra-hightsr}.}
    \label{fig:passivepitch-trf-hightsr}
\end{figure}
As it was the case for the low-$\lambda$ test cases, \cref{fig:passivepitch-trf-hightsr} reports the transfer functions of the torque, thrust, root-bending moment and angular velocity measured from the spectra pairs seen in \cref{fig:passivepitch-spectra-hightsr}.
Unlike the low-$\lambda$ test cases, it can be appreciated that the values of all transfer functions are closer to the neutral value of \SI{0}{\deci\bel} for all values of $f$; moreover, the instability introduced in the angular velocity due to the loads imbalance resulting from the single pitching blade, is visible at $f/\omega = 1$ for the angular velocity transfer function, while it does not appear to affect the loads experienced by the turbine considerably.
Therefore, the load mitigation is due to the alleviation of the oscillations of the blade incidence, despite greater oscillations of the dynamic pressure, confirming that the passive pitch system worked as intended.

\subsection{Discussion} \label{ssec:discussion}
The results presented in this section, and especially those regarding the reduction of the loads standard deviation in \cref{fig:passivepitch-std}, highlight how the passive pitching mechanism is suitable for the goal of reducing the unsteadiness in the loads experienced by the turbine, and therefore reduce the intensity of the fatigue loading on the structure, the drivetrain and the blades, with obvious consequences on the operating life of a tidal turbine and ultimately on its levelised cost of energy.
This is achieved even when the passive pitch implementation is not optimal, as it was in the case of this study: for the experimental setup here tested, both the pitching spring and the double row ball bearings used have generated friction that has proven to be non-negligible on the results achieved by the system, albeit no attempt at modelling this has been carried out.
While the dynamic performance of the passive pitching system is promising, some issues persist on the time-average behaviour of this mechanism: data reported in \cref{fig:passivepitch-meanvalues} shows how complicated it is to reproduce the mean values of thrust and root-bending moment of a canonical turbine equipped with rigid blades, while the torque is seemingly less affected.
This is mostly due to the uncertainty on the actual, time-average pitching angle of the turbine blade when equipped with a passive pitching mechanism: as mentioned in \cref{eq:ode_pitching,eq:ss_pitching}, the actual time-average pitching angle is a function of the equilibrium between the spring-back torque and the hydrodynamic moment generated by the blade.
It can be appreciated that any error on the estimation of the hydrodynamic moment generated by the blade or any error in the actual assembly of the passive pitch system result in different equilibrium locations.
This in turn results in a different time-average pitch of the blade and eventually in different time-average loads on the blade.
Further implementations of the passive pitch system will therefore have to account for the inevitable uncertainty on the hydrodynamic moment generated by the blades, and therefore for a more immediate correction of the time-average pitching angle, if the time-average loads must be kept with a passively pitching blade.

The results here presented show that unsteadiness in the loads generated by the tidal turbine have been reduced by allowing one of the turbine blades to pitch around an axis, with reductions of \SIrange{40}{60}{\percent} in the standard deviation of the thrust, the torque and the angular velocity of the turbine for values of tip-speed ratios for which the turbine generates the highest amount of power; for higher $\lambda$, the reductions are still present albeit with a smaller impact, while for lower $\lambda$ these values might have been overestimated as an artificial component of unsteady loading due to drivetrain instability has been removed.
However, the low-order code that was presented by \cite{Pisetta2022} suggested the loads unsteadiness could be reduced almost completely, especially in the thrust and the root-bending moment, an observation that cannot be made from the experimental results here shown.
Understanding the differences between the low-order code and the experimental reality is then paramount to understand whether the simplified approach can yield realistic results, and thus if this can be used as a design tool for larger-scale tidal turbines.
Two main phenomena that have affected the results here presented were not forecast when developing the low-order code of \cite{Pisetta2022}: these are the friction generated by the bearings and the pitching spring, and the uncertainty in the average pitching angle whose effects on the mean loads was discussed in the last paragraph.
Neither of these effects were present in the simulations of \citet{Dai2022}, which have modelled a similar tidal turbine in a similar inflow.
In this work, the authors have indeed found a reduction in the unsteady thrust generated by the tidal turbine of \SI{75}{\percent} in the worst-case scenario, a value that is compatible with the results of the reduced-order approach.
This suggests that, with an accurate control of the time-average pitch angle and a reduction in the friction that opposes the pitching motion, practical realisations of the passive pitch mechanism will also be able to obtain results in line with those of the reduced-order approach, and be beneficial for the deployment of tidal turbines in realistic scenarios.

In \cref{ssec:ppitch_mech}, a brief rundown of the different sources of friction that affected the mitigation of the blade loads has been given.
It was observed that the bearings were not the only source of friction in the system, as the sliding motion of the pitching spring on the shaft also generates harmful friction: this last component is much harder to estimate than the opposing torque presented by the bearings, as it depends on which fraction of the spring is in contact with the shaft instantaneously, itself a function of the instantaneous torsion of the spring.
As an estimate of the parasitic torsion generated by these sources of friction is necessary to model the whole blade dynamics, the low-order code will necessarily provide optimistic estimates of the blade load reduction.

The consequence of this observation is that any experimental or practical realisation of passive pitch systems must minimise all sources of friction, so to achieve the unsteady load reductions predicted by the low-order codes.
Simultaneously, low-order codes that include an estimate of the friction around the pitching shaft will result in more realistic estimates.

\section{Conclusions} \label{sec:conclusions}
In this study, we have presented a first approach to the implementation of the morphing blade technology initially formulated by \cite{Viola2021,Viola2022p} and \cite{Pisetta2022}: in the experimental campaign carried out, one blade of a three-bladed, speed-controlled model-scale tidal turbine has been allowed to pitch around an axis parallel to its spanwise direction, with the rotation about this axis being constrained by a torsional spring.
The turbine has been subjected to a sheared, turbulent inflow which is representative of the flow full-scale turbines encounter in tidal channels.

From comparison between the loads experienced by the turbine when fitted with canonical rigid blades and when fitted with one passively pitching blade, it has been observed that the passive pitch mechanism has reduced the fluctuations in the torque, the thrust and the root-bending moment generated by the tidal turbine during operation: in particular, this last quantity is reduced by at least \SI{15}{\percent} and up to \SI{45}{\percent}, depending on the tip-speed ratio.
The best test-case for the passive pitch system was that of low tip-speed ratios, as the pitching system mitigated both the amplitude of the loads experienced by the turbine and the unsteadiness in the turbine speed, which the drivetrain speed controller failed to maintain constant. 
Analysis of the power spectra revealed that high performance of the passive pitch system at low tip speed ratios are due to the mitigation of low-speed instability due to the turbine speed controller.
The passive-pitch system performed well at high tip speed ratios as well, where no instability occurred, mitigating load fluctuations across the whole spectrum, and giving a definitive proof of its effectiveness.

While the passive pitching system has indeed been able to achieve a reduction in the unsteady loads experienced by the turbine during its lifetime, some critical observations have to be drawn.
Firstly, the mean loads experienced by the turbine are heavily dependent on the preload of the pitching spring: this can result in higher values of mean thrust and root-bending moment, especially for the case in which the spring preload is higher than needed; this in turn results in larger loads experienced by the turbine structures, which require larger structures thus invalidating the benefits of the passive pitch technology.
Moreover, it was found that the friction generated both by the pitching axis bearing and by the spring as it winds around the pitching shaft is not negligible, having a clear effect on the performance of the passive pitch system.
This information is extremely useful for the implementation of low-order models that aim to simulate the effects of passively pitching blades on the performance of tidal turbines: the introduction of friction around the pitching shaft is straightforward to implement, as this is often estimated by empirical relations that add little computational cost to the models.
It is therefore likely that simple modifications to existing low-order codes have the potential to dramatically improve the accuracy of their results.
Moreover, these results show that the performance of a passive pitch system can be realistically improved to achieve results more similar to those of either CFD simulations \citep{Dai2022} or low-order models \citep{Pisetta2022} by minimising the friction around the pitching shaft.
The results also show that a careful estimate of the preload moment provided by the pitching spring can decrease the thrust and the root-bending moment generated by the turbine without affecting the power generated by the turbine considerably.

The results here presented prove the viability of passive pitch mechanisms in reducing the loads experienced by tidal turbines, an ultimately to reduce their operating costs with minimal effects on the annual power output.
Further attempts at implementing passive pitching mechanisms will have to account for the issues found in this experiment: namely, these will have to ensure that the friction generated by bearings and springs are a negligible fraction of the total pitching moment generated by the blade, as well as ensuring that the preload angle of the pitching spring can more easily and accurately tuned to the desired value.

\section*{CRediT authors contribution statement}
Both SG and GP analysed the data and jointly wrote the first draft of the manuscript.
GP designed and executed the experiments with the help of TD and JS.
IMV conceived and supervised the project.
All authors revised, edited and approved the final version of the manuscript.

\section*{Data disclosure statement}
Data relative to all the results and figures here presented is publicly available at the following address: \url{https://doi.org/10.7488/ds/3483}.

\section*{Declaration of competing interests}
The authors declare that they have no known competing financial interests or personal relationships that could have appeared to influence the work reported in this paper.

\section*{Acknowledgements}
This work was supported by the EPSRC through grant EP/L016680/1, which funded the PhD Scholarship of GP, and grant EP/V009443/1, which was awarded to IMV.

\bibliographystyle{elsarticle-harv}
\bibliography{references,misc}

\appendix

\section{Choice of the preload angle} \label{sec:appendix-preload}
In \cref{ssec:ppitch_mech} we have outlined how the correct value of the preload angle $\beta_\mathrm{pre}$ can be estimated by \cref{eq:betapre}.
Assuming an estimate of the static friction, one can determine the preload angle as
\begin{equation}
    \beta_\mathrm{pre} = \frac{1}{k} (M_h(\beta = 0) + M_\mathrm{fric}(\dot{\beta} = 0)).
\end{equation}
It is sensible to determine the preload angle as a function of the spring constant $k$ and not the opposite, as the reduced-order code shows a clear effect of different values of $k$ on the performance of the passive pitch system \citep{Pisetta2022}.
Some additional constraints on the preload angle are introduced by practical considerations: the actual preload angle that can be guaranteed by the setup is discretised to the spacing in the mounting holes in the casing halves (components F1 and F2 in \cref{fig:pitch-exploded}(a)), which is \SI{10}{\degree} for the setup used here.
This introduces a possible bias between the actual and the desired preload angle of at most \SI{5}{\degree}: the effect of this can be reduced if the preload angle is large enough, so that this error is relatively small.

As the passive pitch system harvests the variations in the hydrodynamic moment generated by gusts, any source of friction on the pitching shaft reduces the torque available to the system and, in principle, hinders its performance.
As previously mentioned, the pitching axis is connected to the turbine hub frame by means of two roller bearings and a spring: both are potential sources of friction that have to be carefully designed.
Using the low-order code of \cite{Pisetta2022}, the time-average pitching moment generated by the blade when operating at $\lambda = 7$ and $U_\infty = \SI{0.8}{\metre\per\second}$ is \SI{1.89}{\newton\metre}, while the fluctuations around this mean have amplitudes that are one order of magnitude smaller.
The motion of the passively pitching blade is expected to be that of an oscillation around the equilibrium location at $\beta = 0$, with a maximum amplitude of \SI{2}{\degree} and a maximum angular velocity about the pitching axis of \SI{36}{\degree\per\second}.
In these conditions, the main source of friction from the ball bearings is the sliding friction between the rotating part and the bearing seal: this generates a moment $M_\mathrm{seal}$ opposing the pitching motion, whose magnitude has been estimated via the empirical formula
\begin{equation}
    M_\mathrm{seal} = K_{S_1} d_{S}^\gamma + K_{S_2},
\end{equation}
where $d_{S}$ is the seal counterface diameter, which is \SI{30}{\milli\meter} for the bearings used in this setup, and $K_{S_1}$, $K_{S_2}$, and $\gamma$ are empirical constants provided by the manufacturer.
The estimated value of $M_\mathrm{seal}$ is \SI{23}{\milli\newton\meter}, approximately \SI{1}{\percent} of the total moment about the pitching axis and \SI{10}{\percent} of the amplitude of the moment oscillations.

\begin{figure}[ht!]
    \centering
    \includegraphics[width=0.5\textwidth]{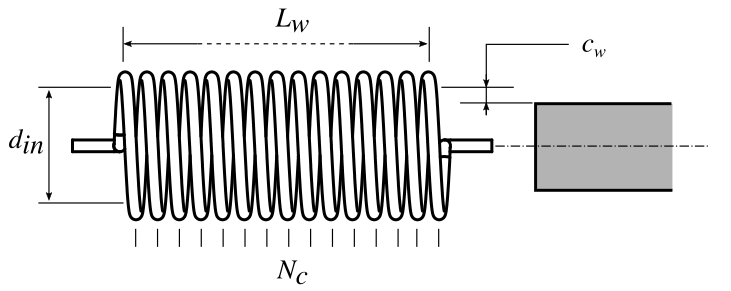}
    \caption{Definition of spring dimension and clearance from the shaft}
    \label{fig:spring-clearance}
\end{figure}
An additional source of friction consists in the contact between the shaft and the spring: a carefully designed setup must account for the reduction in the spring inner diameter as the spring is progressively loaded.
The springs used in this experiment were designed so that these would wrap around the pitching shaft at an angle of \SI{680}{\degree}: this was deemed sufficient as the maximum torsion of the springs would not have exceeded the maximum preload of \SI{550}{\degree} by more than two degrees; the expected inner diameters $d_\mathrm{in}$ and shaft clearances $c_w$ are reported in \cref{tab:spring}.
\begin{table}[ht!]
    \centering
    \caption{Spring diameter and clearance from the shaft when deformed to different torsion angles.}
    \label{tab:spring}
    \vspace{6pt}
    \begin{tabular}{lrrr} \toprule
        Torsion [\si{\degree}] & $N_c$ & $d_\mathrm{in}$ [\si{\milli\meter}] & $c_w$ [\si{\milli\meter}] \\ \midrule
        0 (rest) & 16    & 22.65 & 1.32 \\
        275      & 16.76 & 21.52 & 0.76 \\
        450      & 17.25 & 20.83 & 0.41 \\
        550      & 17.53 & 20.46 & 0.23 \\ \bottomrule
    \end{tabular}
\end{table}
These clearance estimates however assume that the deformation of the spring is such that the spring axis stays parallel to the shaft axis for all possible torsion angles.
When loaded to more than \SI{450}{\degree}, the springs used exhibited a clear bend that might have resulted in contact between the shaft and the spring, thus invalidating the assumption under which the values of $c_w$ reported in \cref{tab:spring} have been obtained.
In fact, during the measurements, it was observed that the performance of the passively pitching blades equipped with springs preloaded by \SI{450}{\degree} or more have not exhibited significant differences from the ones obtained with the rigid blades, hinting that the large spring deflection and the friction it generates markedly impacts the behaviour of the system.

\section{Uncertainty estimation} \label{sec:appendix-uncert}
The half-width of the \SI{95}{\percent} confidence interval on the mean value of a generic measured quantity $Z$ is estimated as 
\begin{equation}
    \varepsilon_{95,Z} = \sqrt{B_Z^2 + (t S_Z)^2},
\end{equation}
where $B_Z$ is the fixed bias in the measurement of $Z$, $t$ is the Student's t-multiplier for \SI{95}{\percent} confidence given the number of degrees of freedom, and $S_Z$ is the standard error on the mean of $Z$.
This last parameter is further defined as
\begin{equation}
    S_Z = \frac{\sqrt{\overline{(Z-\overline{Z})^2}}}{\sqrt{N}},
\end{equation}
where the overline denotes averaging, and $N$ is the number of statistically independent samples of the quantity $Z$.
For the estimation of $t$, the number of degrees of freedom is assumed to be $N-1$.
For the measurements of forces and moments on the turbine, two instantaneous samples are statistically independent if these are separated by more than one turbine rotation period.
Instantaneous current measurements (water temperature and free-current velocity) are instead considered statistically independent if they are separated by more than one convective timescale $D/U_\infty$.
The fixed biases and the confidence intervals of the measured quantities are presented in \cref{tab:unc-measured}.
The confidence intervals reported here are the mean values computed from all acquired timeseries of each of the reported quantity.
\begin{table}[ht!]
    \centering
    \caption{Biases and half-width of the \SI{95}{\percent} confidence intervals on the means of the direct measurements.}
    \label{tab:unc-measured}
    \vspace{6pt}
    \begin{tabular}{llll} \toprule
        Quantity                          & $B$                               & Mean $S$                          & Mean $\varepsilon_{95}$           \\ \midrule
        Temperature $\theta$              & \SI{2e-1}{\kelvin}                & \SI{8e-3}{\kelvin}                & \SI{2e-1}{\kelvin}                \\
        Free-current velocity $U_\infty$  & \SI{8e-3}{\meter\per\second}      & \SI{5e-3}{\meter\per\second}      & \SI{1e-2}{\meter\per\second}      \\
        Turbine angular velocity $\omega$ & \SI{5e-1}{\revolution\per\minute} & \SI{2e-1}{\revolution\per\minute} & \SI{5e-1}{\revolution\per\minute} \\
        Torque $Q$                        & \SI{2e-1}{\newton\meter}          & \SI{2e-1}{\newton\meter}          & \SI{3e-1}{\newton\meter}          \\
        Thrust $T$                        & \SI{4e+0}{\newton}                & \SI{2e+0}{\newton}                & \SI{4e+0}{\newton}                \\
        Root-bending moment $M_y$         & \SI{5e-1}{\newton\meter}          & \SI{3e-1}{\newton\meter}          & \SI{6e-1}{\newton\meter}          \\ \bottomrule
    \end{tabular}
\end{table}

For a generic derived quantity $Y(Z_1, \dots, Z_n)$, function of the measured quantities $Z_1, \dots, Z_n$, the confidence interval is estimated as
\begin{equation}
    \varepsilon_{95,Y} = \sqrt{\sum_{j=1}^{N} \left( \frac{\partial Y}{\partial Z_j} \varepsilon_{95,Z_j} \right)^2}.
\end{equation}
The mean half-widths of the \SI{95}{\percent} confidence intervals are reported in \cref{tab:unc-derived}.

\begin{table}[ht!]
    \centering
    \caption{Half-width of the \SI{95}{\percent} confidence intervals on the means of the derived quantities.}
    \label{tab:unc-derived}
    \vspace{6pt}
    \begin{tabular}{ll} \toprule
        Quantity                              & Mean $\varepsilon_{95}$                \\ \midrule
        Density $\rho$                        &  \SI{4e-2}{\kilo\gram\per\meter\cubed} \\
        Kinematic viscosity $\nu$             &  \SI{5e-6}{\meter\squared\per\second}  \\
		Tip-speed ratio $\lambda$             & \num{8e-2}                             \\
		Torque coefficient $C_Q$              & \num{2e-3}                             \\
		Thrust coefficient $C_T$              & \num{3e-2}                             \\
		Root-bending moment coefficient $C_y$ & \num{6e-3}                             \\ \bottomrule
    \end{tabular}
\end{table}

To estimate the confidence interval on the standard deviation, we instead use a $\chi^2$ approach.
The \SI{95}{\percent} confidence interval on the measured standard deviation $\sigma_Z$ of the timeseries $Z$ is only a function of the number of samples $N$.
The lower and upper bounds $L_\sigma$ and $U_\sigma$ are computed as follows \citep{Sheskin2011}:
\begin{linenomath}
    \begin{align}
        L_\sigma &= \sigma_Z \sqrt{\frac{N-1}{F^{-1}(1-\alpha, N-1)}}, \\
        U_\sigma &= \sigma_Z \sqrt{\frac{N-1}{F^{-1}(\alpha, N-1)}},\\
    \end{align}
\end{linenomath}
where $\alpha = 0.05$ for the \SI{95}{\percent} confidence interval and $F^{-1}$ is the inverse cumulative distribution function of the $\chi^2$ distribution computed at $\alpha$ for $N-1$ degrees of freedom.
With this definition, one has with \SI{95}{\percent} certainty that $\sigma_Z \in [L_\sigma, U_\sigma]$.
For the measurements of temperature and free-current velocity, all timeseries have \num{450} statistically independent samples $N$: for these, the width of the confidence intervals is equivalent to at most \SI{7}{\percent} of the estimated standard deviation.
Measurements of forces, torques, moments and turbine angular velocity are instead acquired with at least \num{290} statistically independent samples: for these, the width of the confidence intervals is equivalent to at most \SI{9}{\percent} of the measured standard deviation.

\end{document}